\documentclass[10pt,oneside,english]{article}

\usepackage[T1]{fontenc}
\usepackage[a4paper]{geometry}
\usepackage[table]{xcolor}
\usepackage[utf8]{inputenc}
\usepackage{algorithm}
\usepackage{algorithmic}
\usepackage{amsfonts}
\usepackage{amsmath}
\usepackage{amsrefs}
\usepackage{amssymb}
\usepackage{amsthm}
\usepackage{authblk}
\usepackage{array}
\usepackage{babel}
\usepackage{booktabs}
\usepackage{calc}
\usepackage{caption}
\usepackage{comment}
\usepackage{enumerate}
\usepackage{fancybox}
\usepackage{float}
\usepackage{graphicx}
\usepackage{hyperref}
\usepackage{mathtools}
\usepackage{multirow}
\usepackage{multicol}
\usepackage{newclude}
\usepackage{rotating}
\usepackage{setspace}
\usepackage{subcaption}
\usepackage{textcomp}
\usepackage{tikz}
\usepackage{url}
\usepackage{xr}

\usepackage{cleveref} 

\makeatletter
\newcommand{\eqnum}{\refstepcounter{equation}\textup{\tagform@{\theequation}}}
\makeatother

\renewcommand{\sec}[1]{Section~\ref{#1}}
\newcommand\publishedin[2][section]{}

\renewcommand\theta\vartheta
\newcommand{\sgn}{\operatorname{sgn}}
\renewcommand\theta\vartheta

\newcommand{\bern}{\mathcal{B}}
\newcommand{\Gauss}{\mathcal{N}}

\newfloat{algorithm}{tbp}{loa}
\floatname{algorithm}{Algorithm}

\newcommand{\W}[0]{\mathbf{W}}
\newcommand{\A}[0]{\mathbf{A}}
\renewcommand{\l}[0]{\mathbf{\ell}}
\newcommand{\Z}[0]{\mathbf{Z}}
\newcommand{\z}[0]{\mathbf{z}}

\renewcommand\epsilon\varepsilon

\newcounter{noqed}

\usetikzlibrary{matrix,arrows,decorations.pathmorphing}
\date{}
\title{ Estimating latent feature-feature interactions \\ in large feature-rich graphs }

\author[1]{Corrado Monti}
\author[1]{Paolo Boldi}
\affil[1]{Dipartimento di Informatica\\
Universit\`a degli Studi di Milano\\
Italy\\
\texttt{\{monti,boldi\}@di.unimi.it}
}

\begin{document}

\maketitle

\begin{abstract}

Complex networks arising in nature are usually modeled as (directed or
undirected) graphs describing some connection between the objects that are
identified with their nodes. In many real-world scenarios, though, those objects
are endowed with properties and attributes (hereby called features). In this
paper, we shall confine our interest to binary features, so that every node has
a precise set of features; we assume that the presence/absence of a link between
two given nodes depends on the features that the two nodes exhibit.

Although the situation described above is truly ubiquitous, there is a limited
body of research dealing with large graphs of this kind. Many previous works
considered homophily as the only possible transmission mechanism translating
node features into links: two nodes will be linked with a probability that
depends on the number of features they share. Other authors, instead, developed
more sophisticated models (often using Bayesian
Networks~\cite{hofman2008bayesian} or Markov Chain Monte Carlo~\cite{MCMC-ML}),
that are indeed able to handle complex feature interactions, but are unfit to
scale to very large networks.

We study a model derived from the works of Miller~\emph{et al.}~\cite{MGJ2009},
where interactions between pairs of features can foster or discourage link
formation. In this work, we will investigate how to estimate the latent
feature-feature interactions in this model. We shall propose two solutions: the
first one assumes feature independence and it is essentially based on a Naive
Bayes approach; the second one consists in using a learning algorithm, which
relaxes the independence  assumption and is based on perceptron-like
techniques. In fact, we show it is possible to cast the model equation in order
to see it as the prediction rule of a perceptron. We analyze how classical
results for the perceptrons can be interpreted in this context; then, we define
a fast and simple perceptron-like algorithm for this task. This approach (that
we call \emph{Llama}, Learning LAtent feature-feature MAtrix) can process
hundreds of millions of links in minutes. Our experiments show that our approach
can be applied even to very large networks.

We then compare these two techniques in two different ways. First we produce
synthetic datasets, obtained by generating random graphs following the model we
adopted. These experiments show how well the Llama algorithm can reconstruct
latent variables in this model. These experiments also provide evidence that the
Naive independence assumptions made by the first approach are detrimental in
practice. Then we consider a real, large-scale citation network where each node
(i.e., paper) can be described by different types of characteristics. This
second set of experiments confirm that our algorithm can find meaningful latent
feature-feature interactions. Furthermore, our framework can be used to assess
how well each set of features can explain the links in the graph.

\end{abstract}

\section{Introduction}
\label{sec:introduction}

The problem of finding a model that describes how complex networks shape their
structure is well studied but still elusive in its full generality. In many
scenarios, though, it is reasonable to assume that the network arises in some
way from a complex interweaving of some features of the nodes. For example, in a
co-authorship network, a link stems more easily between authors with similar
interests; similarly, in a genetic regulatory network, links are affected by the
different biological functions of the regulators.

Many models have been proposed for describing complex network where arcs are
influenced by some features of the nodes. For example, Lattanzi and
Sivakumar~\cite{lattanzi_affiliation} described a model where arcs form at random,
 or as a consequence of shared common features; Caldarelli \emph{et al.}
in~\cite{caldarelli2002scale} proposed a model where arcs are determined by an
arbitrary function of the ``fitness'' of the nodes (i.e., a real-valued property
possessed by each node). More models proposed along this line of research will
be described in~\sec{sec:models-state-of-the-art}.

Although in some cases the relation between features and links is
\emph{homophily} (a link stems more easily between nodes that share a large
portion of \emph{the same} features), we would like to design a model that is
able to capture also more complex behaviors. For example, feature $h$ could
foster links to feature $k$ also when $h \neq k$: e.g., in the case of semantic
relations, a concept tagged as belonging to the category ``Movies'' will often
link to a concept tagged as belonging to ``Directors''. If we consider directed
networks, we would like this relationship between features to be directed:
feature $h$ could foster links towards feature $k$ but not the other way around.
For example, in a citation network, we could easily expect a paper within the
sociology realm to cite a statistics paper, but a link in the opposite direction
will be much harder to find. Finally, some pairs of features could not foster
but rather inhibit link formation: as \emph{``Romeo and Juliet''} narrates,
belonging to rival families could discourage the creation of a link in a
long-term romantic relationship graph.

The theoretical model we are going to describe (based on the work by Miller,
Griffiths and Jordan \cite{MGJ2009}) is able to represent all the aforementioned
kinds of behavior within a unified framework, while at the same time being
simple enough to be computationally useful and scalable, as we will show in the
second part of this work. In this work, we will see how the estimation of the
latent parameters of the model is fundamentally related to perceptron-like
prediction rules, and we will turn this insight into a scalable algorithm able
to extract information also from very large graphs.

In our model a special role is played by the feature-feature matrix $\W$. This
matrix can express the various kinds of interplay between features and links, as
described above; it is a latent, unobservable element of the model, that can
compactly explain the observable links. The question is basically the following:
assuming to know the links of a network and the features every node bears, how
can we estimate how features interact with each other -- i.e., estimating the
matrix $\W$?

This question has a lot of practical implications. Consider for example a
semantic graph~\cite{chein2008graph}, where nodes are concepts, arcs are
semantic relations, and each concept can belong to different categories. Here,
the matrix element $W_{h,k}$ describes how two categories $h$ and $k$ relate to
each other: it summarizes if they interact positively, negatively and how much;
it can therefore be used for measuring the semantic connection between the two
categories. In a linguistic graph (maybe obtained from a large corpus of text),
where a link exists between words used as subjects and those used as objects for
a certain verb, $\W$ describes the semantic areas a given verb can connect. In a
citation network where features are areas of scientific research, the set
$S_k=\{h | W_{h,k} > 0\}$ contains the fields for which the field $k$ is useful,
and so forth.

Many other examples are possible; it is however important to note how many of
these applications require to deal with graphs having a huge number of nodes and
links. We will present concrete examples dealing with tens of millions of nodes.
Operating at this scale demands new techniques; as we will see in
\sec{chap:state-of-the-art}, many of the existing techniques are not able to
scale to this size.

A first idea, that we will describe in \sec{sec:naive}, is to just estimate the
probability of a link from the category pairs we see in the data. We will derive
formally this approach, showing that it can be ascribed to the family of Naive
Bayes learning. In particular, we will see that this estimation requires
independence assumptions that are particularly unrealistic in most practical
cases. For example, consider the semantic link between the entity \emph{``Ronald
Reagan''} and the 1954 Western film \emph{``Cattle Queen of Montana''}; such an
approach will increment, because of the presence of this link, the element of
$\W$ corresponding to $(films,\ U.S.\ presidents)$, regardless of the fact that
this link could already be well explained by $(films,\ actors)$.

Based on the latter observation, we will need to streamline the model: we will
make it deterministic, by fixing its activation function $\phi$. As we will
describe in \sec{sec:llama}, this fact will allow us to see our model equation
as the prediction rule of a perceptron and in the end to develop a more
sophisticated approach based on online machine learning. What we will do is to
see $A$ (the links in the graph) as partially unknown, much like in the link
prediction problem; we will show that, while learning $A$, the internal state of
the perceptron will tend to $\W$. This approach, that we will call
\textsc{Llama} -- Learning LAtent MAtrix -- will overcome the naive assumptions
of the previous model.

In \sec{sec:llama-simulations} we will test this approach on our model, by
simulating graphs obeying the model and then observing how this way of
reconstructing $\W$ behaves. More precisely, we will show how \textsc{Llama} is
able to reconstruct the $\W$ matrix, and how instead the independence
assumptions make the Naive algorithm very far from the goal.

Finally, in \sec{sec:real-data} we will consider a real, large-scale citation
network, where nodes are papers and links represent citations. Here, each node
can be described by different types of characteristics; we will consider
institutions of the authors, and fields of research the paper belongs to. We
will define a notion of \emph{explainability}, a way to measure how a certain
set of features can explain links in a graph according to our framework. Then,
we will prove how real-data experiments confirm that our algorithm can find
meaningful, latent feature-feature interactions from a real network.

\section{Related works}
\label{chap:state-of-the-art}
\label{sec:models-state-of-the-art}

The interplay between features and links in a network was investigated
separately in different fields. Indeed, interpreting links as a result of
features of each node has in fact a solid empirical background. For example, the
dualism between ``persons and groups'' as an underlying mechanism for social
connections was first investigated by Breiger~\cite{breiger} in 1974. Within
sociology, the simple phenomenon of homophily -- ``similarity breeds
connection'' -- received a great deal of attention:  McPherson~\emph{et
al.}~\cite{mcpherson2001birds} presented evidence and investigated on the role
of homophily in social ties; considered features included race and ethnicity,
social status, and geographical location. Bisgin~\emph{et
al.}~\cite{homophilymedia} studied instead the role of interests in online
social media (specifically, Last.fm, LiveJournal, and BlogCatalog), finding
however that the role of interests as features is weak on those online
networks---at least when considering homophily only.

In some fields, behaviors more complex than homophily were considered as well.
Tendencies of such kind, where nodes with certain features tend to connect to
other types of nodes, are called \emph{mixing patterns} in sociology and are
often described by a matrix, where the element $(i,j)$ describes the
relationship between a feature $i$ and a feature $j$. In epidemiology, mixing
patterns have proven to be greatly beneficial in analyzing the spread of
contagions. For example, they appeared to be a crucial factor in tracking the
spread of sexual diseases~\cite{aral1999sexual} as well as in modeling the
transmission of respiratory infections~\cite{mossong2008social}. For this
reason, such matrices are also called ``Who Acquires Infection From Whom''
(WAIFW) matrices, and have been empirically assessed in the
field~\cite{hens2009mining,isella2011close}. In biology and bioinformatics, a
seminal study by Menche \emph{et al.}~\cite{menche2015uncovering} highlighted
the connections between the interactome (the network of the physical and
metabolical interactions within a cell) and the diseases each component was
associated with, observing a clustering of disease-associated proteins.

The empirical evidence presented in various fields, combined with the existence
of large datasets available in the web, and the increase of computational
resources, fostered some investigation of models of graph endowed with features.

\paragraph{Class models.}

A popular framework has been that of \emph{latent class models}: in these
models, every node belongs to exactly one class, and this class influences the
links it may be involved into. The \mbox{best-known} example is the stochastic
block model~\cite{nowicki2001estimation, blockmodel}: in this model, it is
assumed that each pair of classes has a certain probability of determining a
link, and Snijders and Nowicki~\cite{blockmodel} study how to infer those
probabilities; they also investigate how to determine the class assignments,
leading to a sort of community detection algorithm. Hofman and Wiggins
\cite{hofman2008bayesian} devised a variant of this scheme, by specifying only
within-class probabilities and between-class probabilities. Another useful
adaptation involves sharing only the between-class probability and specifying
instead the within-class probabilities separately for each class, allowing to
characterize each with a certain degree of homophily. Both these approaches
exemplify the need to reduce the number of parameters of the original block
model, in order to facilitate the estimation of its parameters. Kemp \emph{et
al.}~\cite{kemp2006learning}, and Xu~\emph{et al.}~\cite{xu2006learning},
studied and applied a non-parametric generalization of the model which allows
for an infinite number of classes (therefore called \emph{infinite relational
model}). It permits application on data where the information about class is not
provided directly. They use a Gibbs sampling technique to infer model
parameters.

A well-known shortcoming of the class-based models is the proliferation of
classes~\cite{MGJ2009}, since dividing a class according to a new feature leads
to two different classes: if we have a class for ``students'' and then we wish
to account for the gender too, we will have to split that class in ``female
students'' and ``male students''. This approach is impractical and in many cases
it leads to overlook significant dynamics. In order to overcome this limitation,
some authors \cite{mixed-membership} extended classical class-based models to
allow mixed membership. Here, the model of classes remains, but with a fuzzy
approach: each node can be ``split'' among multiple classes, and in practice
class assignments become represented by a probability distribution.

\paragraph{Feature models.}

Contrary to class-based models, \emph{feature-based models} propose a more
natural approach for nodes with multiple attributes: in those models, each node
is endowed with a whole vector of features. Therefore, feature-based models can
be seen as a generalization of class-based models: in fact, when all the
vectors have exactly one non-zero component, the model has the same expressive
power of class-based ones. Features can be real-valued -- as
in~\cite{latent-factor-models} -- or binary, where the set of nodes exhibiting a
feature is crisp, and not fuzzy, like in~\cite{meeds2006modeling}.

Many works in this direction proposed models that only allow for homophily,
forbidding any other interaction among features. A seminal example is that of
\emph{affiliation networks}~\cite{lattanzi_affiliation} by Lattanzi and
Sivakumar; in that work, a social graph is produced by a latent bipartite
network of \emph{actors} and \emph{societies}; links among actors are fostered
by a connection to the same society. Gong \emph{et al} \cite{gong2012evolution}
analyzed a real feature-rich social network -- Google+ -- through a generative,
feature-based network model based on homophily.

Our attention will focus instead on models able to grasp more complex
behavior than homophily, following the aforementioned empirical evidence from
social networks, epidemiology and bioinformatics.

\paragraph{MAG model family.}

Within this stream of research, an important line of work has been explored by
Kim, Leskovec and others~\cite{leskovec-mag}, under the name of
\emph{multiplicative attribute graphs}. There, every feature is described by a
two-by-two matrix, with real-valued elements. Those elements describe the
probabilities of the creation of a link in all the possible cases of that
feature appearing or not appearing on a given pair of nodes. As a consequence,
it can be thought as a feature-rich special case of their previous Kronecker
model~\cite{leskovec2010kronecker}. This model has been further extended to
include many other factors; notably, they have modified it to be
dynamic~\cite{kim2013nonparametric}: features can be born and die, and only
alive features bear effects. However, the complexity of this model prevents it
from being used on large-scale networks. The same authors have
proposed~\cite{kim2011modeling} an expectation-maximization algorithm to
estimate the parameters of their base model; nonetheless, reported experiments
are on graphs with thousands of nodes at most. In the dynamic version, they
report examples on hundreds of nodes (e.g., they find that by fitting the
interactions of characters in a Lord of the Ring movie, their features
effectively model the different subplots). In this work, instead, we wish to
handle networks of much larger size: in the experimental part, we will show
examples with many millions of nodes, for which we are able to estimate model
parameters very efficiently.

\paragraph{MGJ model family.}

In 2009, Miller, Griffiths and Jordan~\cite{MGJ2009} proposed a feature-based
model to describe the link probability between two nodes by considering
interactions between all the pairs of features of the two nodes. They show how
by inferring features and their interactions on small graphs (hundreds of
nodes), they are able to predict links with a very high accuracy (measured
through the area under the ROC curve). The estimation technique they propose is
not exact (since this would be intractable~\cite{GG06}), but it is based on a
Markov Chain Monte Carlo (MCMC) method~\cite{MCMC-ML}.

Their model can be interpreted as a generative model; they chose, however, not
to investigate its structural properties in terms of the resulting network
structure. Subsequent work~\cite{pfeiffer2014attributed} focused on this goal,
being able to generate feature-rich graphs with realistic features, but they did
not try to estimate the latent variables of the model necessary to predict
links. In this work, we will build on the evidence gained in our previous
work~\cite{bolcrimon}, that shows how the Miller-Griffiths-Jordan model (further
extended to exhibit competition dynamics in feature generation) can be a
powerful tool to generate networks with realistic, global properties (e.g.
distance distribution, degree distribution, fraction of reachable pairs, etc.).
As explained in previous work~\cite{bolcrimon}, this model can at the same time
be used to synthesize realistic graphs by itself, or as a way to generate, given
a real graph, a different one with similar characteristics.

Despite the capabilities of the MGJ model~\cite{MGJ2009}, however, the choice of
using a MCMC technique in the original work~\cite{MCMC-ML} revealed itself
inadequate to work on datasets larger than some hundreds of nodes. As noted by
Griffiths \emph{et al.} in 2010~\cite{GG06}, there is a need for computationally
efficient models as well as reliable inference schemes for modeling multiple
memberships. Menon and Elkan~\cite{menon2011link} noted how the inadequacy in
handling large graphs underpinned this work, and many similar ones, and ascribed
this flaw to the MCMC method.

There has been, since, a certain amount of studies on how to apply the MGJ model
on larger graphs. The two aforementioned works, for example, tried to solve this
problem in different ways. Griffiths and Ghahramani~\cite{GG06} described a
simpler model: they removed from the original model~\cite{MGJ2009} the
possibility of having negative interaction between features; also, they fixed
the activation function of the model (a component which we will carefully
explain in the next section); in this way, they obtained a framework that is
more computationally efficient, and can be applied to graphs of up to ten
thousand nodes.

Menon and Elkan~\cite{menon2011link}, instead, slightly enriched the model, by
introducing a bias term for each node; then, they proposed a new estimation
technique, based on stochastic gradient descent. A main focus there was to avoid
undersampling non-links to overcome class imbalance, since despite it being
``the standard strategy to overcome imbalance in supervised learning'' it has
the ``disadvantage of necessarily throw[ing] out information in the training
set''. To overcome this problem (that we solve instead in the standard way of
undersampling, see \cref{chap:inferring-w}), they design a sophisticated
approach centered on the direct optimization of the area under the ROC
curve.\footnote{Recent works~\cite{yang2015evaluating} have indicated
empirically as well as theoretically how employing this measure in link
prediction leads to severely misleading results.} With this technique, they can
handle graphs with thousands of nodes. A different approach, that obtained
similar results on graphs of the same size, is~\cite{doppa2010learning}, where
the authors propose an SVM-based estimation of parameters, and report it being
able to run on graphs as large as two thousands nodes in 42 minutes. Our
approach runs in around 15 minutes on graphs three orders of magnitude bigger.

The task we are defining ultimately falls into the realm of latent variable
models~\cite{everett2013introduction}, since we are trying to explain a set of
manifest variables -- links and features -- through a set of latent variables --
the feature-feature interaction weights, i.e., the elements $W_{h,k}$ of the
matrix $\W$. If, like in our case, manifest variables are categorical, we
usually talk about \emph{latent structure models}, that have been studied as
such by statisticians and social scientist since the 1950's. Lazarsfeld started
studying the statistics behind these models in an effort to explain people
answers to psychological questions (specifically,
in~\cite{stouffer1950measurement}, answers from World War II soldiers) through
quantifiable, latent traits~\cite{lazarsfeld1959latent}. These techniques were
improved by later studies~\cite{goodman1974exploratory, henry1983latent};
however, these techniques---conceived for traditional social studies---were
designed for small groups; the use cases described there do not usually involve
more than a hundred of nodes. We require our techniques to work with millions of
nodes, and hundreds of millions of links.

Previous literature has also treated linked document corpora, where features are
the words contained in each document (e.g.,~\cite{liu2009topic}
and~\cite{chang2009relational}). In these works, authors build a link prediction
model obtained from LDA, that considers both links and features of each node.
However, the largest graphs considered in these works have about $10^3$ nodes
(with $\sim10^4$ possible features), and they do not provide the time required
to train the model. \cite{henderson2009applying}~developed an LDA approach
explicitly tailored for ``large graphs'' --- but without any external feature
information for nodes: they rather reconstruct this external information from
scratch; the largest graph they considered has about $10^4$ nodes and $10^5$
links, for which they report a running time of $45-60$ minutes.

In this work, we too will employ a model of the Miller-Griffiths-Jordan  family,
that we will thoroughly describe in~\sec{chap:models}. As mentioned above, a way
to generate realistic graphs with this model was studied in~\cite{bolcrimon}.
Here, we will propose some further considerations on that model that will lead
(in \cref{chap:inferring-w}) to various techniques aimed at estimating the main
parameter of the model, i.e., the feature-feature matrix. We will test those
methods on synthetic data generated by our model in
\cref{sec:llama-simulations}. In \cref{sec:real-data} we will try our methods
empirically on real networks whose size is unmatched by previous literature.

\section{Our framework}
\label{chap:models}
\label{sec:graph-model}

Let us briefly present the main actors in our theoretical framework. In this
work, we will treat the following objects as (at least partially) observable:

\begin{itemize}

    \item The (possibly directed) graph $G=(N,A)$, where $N$ is the set of $n$
    nodes, whereas $A \subseteq N \times N$ is the set of links; for
    the sake of simplicity, in this work we assume that self-loops (i.e., links
    of the form $(i,i)$) are allowed.

    \item A set $F$ of $m$ features.

    \item A node-feature association $Z \subseteq N \times F$.

\end{itemize}

We will denote these objects through their matrix-equivalent representation.
More precisely, $G=(N,A)$ will be represented as a matrix $\mathbf{A} \in \{ 0,1
\} ^ {n \times n}$ (fixing some arbitrary ordering on the nodes); $Z$ will be
represented as a matrix $\Z \in \{ 0,1 \} ^ {n \times m}$ (again fixing some
arbitrary ordering on the features).
In the following, $A_{i,j}$ will refer to the element in the $i$-th row and
$j$-th column of the matrix $\mathbf{A}$.

The -- typically unobservable -- objects that will define our network model will
be the following: \begin{itemize}

    \item A matrix $\mathbf{W} \in \mathbb{R}^{m \times m}$, that represent how
    features interact with each other. The idea is that a high value for
    $W_{h,k}$ means that the presence of feature $h$ in a node $i$ and of
    feature $k$ in node $j$ will foster the creation of a link from $i$ to $j$.
    Conversely, a negative value will indicate that such a link will be
    inhibited by $h$ and $k$. Naturally, the magnitude of $\big| W_{h,k} \big|$
    will determine the force of these effects.

    We will refer to $\mathbf{W}$ as the \emph{latent feature-feature matrix}.

    \item A monotonically increasing function $\phi : \mathbb{R} \rightarrow
    [0,1]$ that will assign a probability to a link $(i,j)$, given the real
    number resulting from applying $\mathbf{W}$ to the features of $i$ and $j$;
    we will call such a function our \emph{activation function}, in analogy with
    neural networks~\cite{hertz1991introduction}.

\end{itemize}

The relationship between those actors is described formally by the following
equation, that fully defines our model:

\begin{equation}
    \label{eq:basic-model}
    \mathbb{P} \Big( (i,j) \in A \Big) = \phi \bigg(
        \sum_{h} \sum_{k} Z_{i,h} W_{h,k} Z_{j,k}
    \bigg)
\end{equation}

In other words, the probability of a link is higher when the sum of $W_{h,k}$ is
higher, where $h,k$ are all the (ordered) pairs of features appearing in the
considered pair of nodes. We will now carefully detail this equation in the
following sections.

\subsection{Model parameters}
\label{sec:model-parameters}

\paragraph{Analysis of the latent feature-feature matrix.}

Let us point out how different choices for $\W$ can lead to many different
kinds of interplay between links and features. The simplest case is $\W=I$ (the
identity matrix). Since its only non-zero elements are those of the form
$(k,k)$, the only non-zero elements in the summation are those with
$Z_{i,k}=Z_{j,k}=1$. Therefore, the behavior of the model in this case is that
of pure \emph{homophily}: the more features in common, the higher the probability
of a link (remember that $\phi$ is monotonic).

More generally, as we said, a positive entry $W_{h,k}>0$ will
indicate a positive influence on the formation of a link from nodes with feature
$h$ to nodes with feature $k$. In the special case of an
undirected graph, we will have a symmetric matrix -- that is, $W_{h,k}=W_{k,h}$
for all $h$ and $k$.

$\W$ can be used to express also other behaviors. If $\sum_{k}{W_{h,k}}$ is
large, this fact will indicate that nodes with feature $h$ will be highly
connected -- specifically, they will have a large number of out-links. A large sum for a
column of $\W$, that is a large value for $\sum_{k}{W_{k,h}}$, will imply,
in turn, that nodes with feature $h$ to have many in-links.

\paragraph{Choice of the activation function.}

\begin{figure}
    \includegraphics[width=0.9\columnwidth]{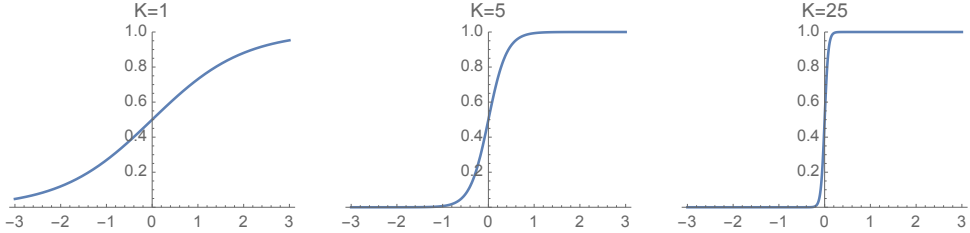}
    \caption{\label{fig:sigmoids}
        A sigmoid activation function $\phi$, with different choices
        for $K$. $K$ regulates its smoothness, and for $K \rightarrow \infty$ it
        approaches a step function.
    }
\end{figure}

The activation function $\phi$ will determine how the real numbers resulting
from $\sum_{h} \sum_{k} Z_{i,h} W_{h,k} Z_{j,k}$ will be translated into
a probability for the event $\big\{ (i,j) \in A \big\}$.
Since we require $\phi$ to be monotonically increasing, its role is just to
\emph{shape} the resulting distribution.

Throughout this work, and following previous literature~\cite{MGJ2009}, we will
focus on activation functions that can be expressed as a sigmoid:

\begin{equation}
    \label{eq:sigmoid}
    \phi(x) = \big( e^{K(\theta-x)}+1 \big) ^{-1}
\end{equation}

The parameter $\theta \in \mathbb{R}$ is the center of the sigmoid, whereas $K
\in (0,\infty)$ regulates its smoothness. Figure~\ref{fig:sigmoids} depicts how $K$ influences
the resulting probabilities (when $\theta=0$). We will look at both these
quantities as \emph{a priori} parameters of the model. We will also extend the
domain of $K$ to the special value $K=\infty$, for which $\phi$ is the step
function\footnote{We will use the notation \begin{equation*}
    \chi_I(x) =
    \begin{cases}
      1 & \text{if $x \in I$} \\
      0 & \text{if $x \notin I$}. \\
    \end{cases}
\end{equation*}} $\chi_{(\theta,\infty)}$. Letting $K=\infty$ will make our
model fully deterministic---all the probabilities become either $1$ or $0$. We will
see how this simplification can turn our model into an important framework for
mining information from a complex network.

\subsection{An algebraic point of view}

For some applications, it will be useful to consider the model expressed by
\eqref{eq:basic-model} as a matrix operation. As introduced in the previous
section, $\Z$ is the $n \times m$ node-feature indicator matrix.

With this notation, we can express \eqref{eq:basic-model} as

\begin{equation}
    \label{eq:matrix-view}
    \mathbf{P} = \phi \Big( \Z \W \Z^{T} \Big)
\end{equation}
where $\phi$ here denotes the natural element-wise generalization of our
activation function --- i.e., it simply applies it to all the elements of the
matrix. The resulting matrix $\mathbf{P}$ is a matrix that describes the
probabilities of $\A$: that is, its element $P_{i,j}$ defines the probability
that $L_{i,j}=1$ or equivalently that $(i,j) \in A$.  You can think of $P$ as an
uncertain graph~\cite{khan2015uncertain, potamias2010k}, of which $A$ is a
realization (sometimes called a \emph{world}~\cite{dalvi2007efficient}).
Uncertain graphs are a convenient representation of graph distributions, in the
same spirit as the classical Erd\H{o}s-R\'enyi model: in an uncertain graph the
node set is fixed and each arc has a certain probability of being present (arcs
are independent from one another). Many useful statistical properties of the
graph distribution associated to an uncertain graph (e.g., the expected number
of connected components) can be connected to properties of the uncertain graph
itself, seen as a simple weighted graph; it is this connection that made
uncertain graphs particularly popular in some contexts.

While this view is simple and concise, it may be of little use from a
computational perspective. In concrete applications $n$ will be very large;
also, algorithms that could be of use in dealing \emph{directly} with this
representation do not run in linear time---the most notable example being
matrix factorization (e.g., computing the SVD~\cite{trefethen1997numerical}).

It is useful, however, to view \eqref{eq:matrix-view} separately for each row
of the matrix. In practice, this means computing the set of out-links of a
single node. This operation allows us to treat a single node at a time,
permitting the design of \emph{online} algorithms, requiring a single pass on
all the nodes.

Moreover, this interpretation renders $\W$ a (possibly asymmetric) similarity
function: if we represent nodes $i$ and $j$ through their corresponding rows in
$\Z$ (indicating them as $\mathbf{z}_i$ and $\mathbf{z}_j$) then our
feature-feature matrix can be seen as a function that given these two vectors
computes a real number representing a weight for the pair $(i,j)$. In the
special case of $\W=I$ this is the standard inner product $\langle \mathbf{z}_i,
\mathbf{z}_j \rangle$; in this case the similarity of those two vectors is just
the number of features they share, thus implementing homophily. Instead, for a
general $\W$ this similarity is $\langle \mathbf{z}_i, \mathbf{z}_j \rangle_\W$
(although $\W$ is not necessarily symmetric or positive definite). In this sense
$\W$ can be seen as a function $W: 2^F \times 2^F \rightarrow \mathbb{R}$, that
acts as a \emph{kernel} for sets of features.

\subsection{Intrinsic dimensionality and explainability}
\label{sec:intrinsic-dim}

Every fixed graph $G$ has a probability that depends on the feature-feature
matrix and, of course, on $Z$, that is, on the choice of the features that we
associate with every node, and ultimately on the set of features we choose.

Some sets of features will make the graph more probable than others; we might
then say that the \emph{explainability} is a property of the chosen set of
features for a certain graph $G$. We will measure it in practice in some
scenarios in the third, experimental, part of this work.

For the moment, let us point out that the number of features can be seen
as an \emph{intrinsic dimensionality} of the graph $G$: if the graph could be
explained by our model without any error at all, then the same information of
$G$ is in fact contained in $\Z$ and $\W$. In that case, we might say that the
out-links of node $i$ (described in the graph by $\l_i$, the $i$-th row of
the adjacency matrix $\A$) could be equally represented by $\mathbf{z}_i$, thus
with a much smaller dimension: specifically, with a vector of $m$ elements.

In fact, $n$ is a natural upper bound for $m$. Let us use the nodes themselves
as features (i.e., $F=V$), associating with every node $i$ the only feature $i$
(i.e, setting $\Z=I$). If $\W=A$ then the graph
will be always perfectly explained: it would be enough to choose $\phi$ as the
step function $\chi_{(0,\infty)}$ to make the results of our model
identical to the graph, since

\begin{equation*}
    \mathbb{P} \Big( (i,j) \in A \Big) = P_{i,j} =
    \left[\phi\left(\Z\W\Z^T\right)\right]_{i,j}=\phi(W_{i,j})=
  \begin{cases}
      1 &\text{ if $(i,j) \in A$} \\
      0 &\text{ otherwise.}
  \end{cases}
\end{equation*}

Naturally, this choice of features does not tell us much; in practice, we
obviously want $m \ll n$. For this, we allow for the introduction of some degree
of approximation; some links will be wrongly predicted by our model, because it
will expect their categories to link to each other. We shall call this effect
\emph{generalization error}. We will see in experiments how it can be measured
and how it is intimately connected with the explainability of a set of features
in a graph.

\subsection{Introducing normalization}

Let us now present some interesting variants of the proposed model. In many
real-world scenarios, we can speculate that not all features are created equal.
For example, in the formation of a friendship link between two people, discovering
that they both have watched a very popular movie may not give us much insight;
knowing instead that they both have seen an underground movie that few people
have appreciated could give to their friendship link a more solid background. In
other words, in some cases rarest features matter more.

\paragraph{Column normalization.}

To implement this effect, we can normalize $\Z$ by column (recall that
columns correspond to features) in our equation, defining

\begin{equation*}
    \overleftarrow{Z}_{i,h} = \frac{Z_{i,h}}{||{Z_{-,h}}||_p}
\end{equation*}
where $Z_{-,h}$ denotes the $h$-th column of $\Z$ and $||-||_p$ represents the
$\ell^p$ norm, for some chosen $p$. The notation $\overleftarrow{\Z}$ is used to
emphasize the fact that, if $p=1$, this normalization yields a left-stochastic
(i.e., column-stochastic) matrix. Each column can be seen in this case as a probability distribution among nodes, uniform on nodes having that feature and null on the
others. If we plug $\overleftarrow{\Z}$ in place of $\Z$ in
\eqref{eq:basic-model}, we obtain
\begin{equation}
    \mathbb{P} \Big( (i,j) \in A \Big) =
    \phi \bigg(
        \sum_{h} \sum_{k} \overleftarrow{Z}_{i,h} W_{h,k} \overleftarrow{Z}_{j,k}
    \bigg) =
    \phi \bigg(
        \sum_{h} \sum_{k} \frac
            { Z_{i,h} W_{h,k} Z_{j,k} }
            { ||Z_{-,h}||_p\cdot ||Z_{-,k}||_p }
    \bigg)
\end{equation}
thus reaching the effect we wanted: inside the summation, rare features will
bear more weight, and common features will be of lesser importance. This can
also be seen as an adaptation of a tf-idf-like schema~\cite{wu2008interpreting}
to our context.

\paragraph{Row normalization.}
\label{par:row-normalization}

In other contexts, row normalizations might be desirable instead. The fact that
two people $x$ and $y$ are friends of the same individual $z$ in Facebook may be
a sign indicating that they have some common interest, and that they may become
friends in the future; however, if $x$ is a public figure then the fact that he
is friend with $z$ is not really significant, and does not tell us much about
possible future friendship with $y$. In other words, nodes with few features may
matter more.

Formally, row normalization is defined as
\begin{equation*}
    \overrightarrow{Z}_{i,h} = \frac{Z_{i,h}}{||Z_{i,-}||_p}.
\end{equation*}
where $Z_{i,-}$ is the $i$-th row of $\Z$.
Again, we used the notation $\overrightarrow{\Z}$ because when $p=1$ we obtain
a right-stochastic matrix.

\section{Inferring feature-feature interaction}

\label{chap:inferring-w}

The fundamental agent in shaping the graph in our framework is, as stated in the
previous section, the feature-feature matrix $\W$. In many applications,
however, the information represented by $\W$ is not directly available: in a
social network, we can observe friendship links and characteristics of each
person, but the relationship between the characteristics is latent and not
observable. This is the case for many other scenarios: in a linked document
corpora where documents are described by a set of topics, we do not know how
different topics foster or discourage links. Knowing (at least partially) links
and features of each node, but ignoring how features interact with each other,
is also a common trait of all the examples we mentioned before.

As discussed in \sec{sec:introduction}, knowing the latent feature-feature
matrix has a lot of practical implications: it can summarize effectively how
features interact with each other -- in the case of a semantic network tagged
with categories, it means getting a hold of which categories are semantically
connected, for a citation network it means being able to identify which fields of
research are being useful for a certain field, and so on. More generally, as we
discussed in \sec{sec:graph-model}, knowing $\W$ means being able to represent
all the information expressed by the graph in a more succinct way.

The problem we wish to solve is therefore the following: assuming to know $A$
and $\Z$, how can we reconstruct a plausible $\W$? In other words: if we know
the arcs in a graph, and each node is characterized by a set of (binary)
features, how can we estimate how features interact with each other?

\subsection{A naive approach}
\label{sec:naive}

\newcommand{\ZE}{\mathcal{Z}}

Let us first describe a naive approach to construe the latent feature-feature
matrix $\W$; remember that we are assuming~(\ref{eq:basic-model}),
where $Z$, $A$ and $\phi$ are fixed (the role of $\phi$ will be discussed
below) and we aim at choosing $\W$ as to maximize the probability of $A$.

More precisely, we shall use a naive Bayes
technique~\cite{BishopPatternRec}, estimating the probability of existence of a link through maximum likelihood and assuming independence between features; that is, we are going to assume that the
events $\{Z_{i,h} = 1\}$ and $\{Z_{i,k} = 1\}$ are independent for $h$ and $k$.

Let us introduce the following notation:

\begin{itemize}

    \item let $N_k \subseteq N$ be the set of nodes with the feature $k$, i.e.
    $N_k = \{i \in N | Z_{i,k} = 1 \}$;

    \item conversely, let us write $F_i$ for the set of features sported by a
    node $i$, that is \\ $F_i = \{ k \in F | Z_{i,k} = 1\}$;

    \item let us also use $\ZE_{i,k}$ to denote the event $\{ Z_{i,k} = 1 \}$.

\end{itemize}

Now, fixing two features $h$ and $k$, let us consider the probability $p_{h,k}$
that there is a link between two arbitrary nodes with those features, such as $i
\in N_h$ and $j \in N_k$:

\begin{equation*}
	p_{h,k} := \mathbb{P} \Big( (i,j) \in A \Big| \ZE_{i,h} \cap \ZE_{j, k} \Big).
\end{equation*}
Said otherwise, $p_{h,k}$ represents the probability that two nodes $(i,j)$
happen to be connected, if we assume that $i$ has feature $h$ and $j$ has
feature $k$.
This quantity can be estimated as the fraction of pairs $(i,j)$ such that both $\ZE_{i,h} $ and $ \ZE_{j, k}$ are true, that happen to be links. In other
words,
\begin{equation*}
	p_{h,k} = \frac{|(N_h \times N_k) \cap A|}{|N_h|\cdot |N_k|}
\end{equation*}

Here, and in the following, we are assuming that self-loops are allowed.
For a specific pair of nodes $(i,j)$, the probability of the presence of a
link under the full knowledge of $\Z$ is given by

\begin{equation*}
    \mathbb{P} \Big( (i,j) \in A \Big|
        \big( \bigcap_{h \in F_i} \ZE_{i,h} \big) \cap
        \big( \bigcap_{h \in F_j} \ZE_{j,h} \big)
    \Big).
\end{equation*}
This is the probability that $(i,j)$ are connected, given that we know their
common features. Let us naively assume that $\ZE_{i,h}$ and $\ZE_{j,k}$ are
independent for all $i,j,h,k$ with $i \neq j$ and $h \neq k$; we also assume that they are
independent even under the knowledge that $(i,j) \in A$. Then, under these naive
independence assumptions, the last probability can be expressed as
\begin{equation*}
    \prod_{h \in F_i} \prod_{k \in F_j} \mathbb{P} \Big(
        (i,j) \in A \Big| \ZE_{i,h} \cap \ZE_{j, k}
    \Big) = \prod_{h \in F_i} \prod_{k \in F_j} p_{h,k}
\end{equation*}

Let us define $\W$ as:

\begin{equation}
	W_{h,k}    =   \log \frac
                        {   |(N_h \times N_k) \cap A|}
                        {|N_h| \cdot |N_k|}
    \label{eq:naive}
\end{equation}

We will now check that such a matrix is correct. Considering again the
definition of our model \eqref{eq:basic-model} and plugging in the matrix $\W$
just defined, we obtain:

\begin{multline}
    \mathbb{P} \Big( (i,j) \in A \Big) = \phi \bigg(
        \sum_{h \in F_i} \sum_{k \in F_j} W_{h,k}
    \bigg) =
    \phi \Big(  \log \prod_{h \in F_i} \prod_{k \in F_j}  \frac
                        {   |(N_h \times N_k) \cap A|}
                        {|N_h| \cdot |N_k|} \Big) = \\
    = \phi \Big( \log \prod_{h \in F_i} \prod_{k \in F_j} p_{h,k} \Big)
    = \phi \Big( \log \mathbb{P} \big( (i,j) \in A \big|
    \big( \bigcap_{h \in F_i} \ZE_{i,h} \big) \cap
    \big( \bigcap_{h \in F_j} \ZE_{j,h} \big) \big) \Big) = \\
    = \phi \Big( \log \mathbb{P} \big( (i,j) \in A \big| \Z \big) \Big)
    \label{eq:naive-proof}
\end{multline}
This fact confirms that, for a certain choice of $\phi$ (namely\footnote{We need
the $\min$ in this formula to respect our assumption that $\phi$ only has values in
$[0,1]$. However, it does not change anything in practice, since in
\eqref{eq:naive-proof} the argument of the logarithm is a probability, and
therefore it is forced to be in $[0,1]$. }, $\phi(x) = \min(1, e^x)$ ) and under
the previously mentioned independence assumptions, this estimate of $\W$ is
correct for our model.

\subsection{A perceptron-based approach}
\label{sec:llama}

The independence assumptions behind the naive approach (hereby referred to as
\textsc{Naive}) are not realistic. One of the potentially undesirable
consequences of such assumptions is that the responsibility for the existence of
a link are shared among all the features of the two involved entities. To
understand how misleading this approach can be, consider a semantic link between
the entity for ``Ronald Reagan'' and the one for \emph{``Cattle Queen of
Montana''} (a 1954 Western film starring Ronald Reagan). \textsc{Naive} will
count that link as a member of the set $\{(N_{presidents} \times N_{movies})
\cap A\}$, and it will consequently increase the corresponding entry
($W_{presidents,\ movies}$) in the feature-feature matrix. We would like instead
to design an algorithm that is able to recognize that this link is already well
explained by the matrix element $W_{actors,\ movies}$ and that does not enforce
a false association between politicians and Western movies. In other words, we
want an algorithm that perceives if some feature is already explaining a link,
and updates its estimate of $\W$ only if it is not. In this perspective, we want
to properly cast our problem in the setting of machine learning.

\paragraph{A deterministic model.}

In order to obtain this result, let us simplify our framework by letting
$\phi$ be the step function $\chi_{(0,\infty)}$, that is
\begin{equation*}
    \phi(x) = \begin{cases}
      1 & \mbox{if } x > 0, \\
      0 & \mbox{otherwise.} \\
    \end{cases}
\end{equation*}
This can be also seen as a sigmoid \eqref{eq:sigmoid} whose parameters are
$\theta = 0$ and $K \rightarrow \infty$. In previous work~\cite{bolcrimon}, we
found that, even if such an activation function produces a more disconnected
network, the network degree distribution will converge even more sharply to a
power law.

It is important to note that this choice will make our model fully
deterministic. In other words, given the complete knowledge of $\Z$ and $\W$,
the model will not allow for any missing or wrong link. For this
reason, with this model we can not measure the \emph{likelihood} of a real
network; instead, we will just separate its links into \emph{explained} and
\emph{unexplained} by the model with respect to a certain set of feature $F$.

\paragraph{A decision rule.}
By using this deterministic activation function, the equation of our model
\eqref{eq:basic-model} becomes:

\begin{equation}
    \label{eq:det-model}
    (i,j) \in A \iff \sum_{h} \sum_{k} Z_{i,h} Z_{j,k} W_{h,k} > 0
\end{equation}
Let us indicate the $i$-th row of $\Z$ with $\z_i$ (as a column vector), the
outer product with $\otimes$ and the Hadamard product with $\circ$. Then, we
can alternatively write the above rule in one of these two equivalent forms:


\begin{equation*}
    (i,j) \in A \iff \z_i^T \W \z_j > 0
\end{equation*}
or
\begin{equation}
    \label{eq:decision-rule}
    (i,j) \in A \iff \sum_{h,k} \Big[ \big( \z_i \otimes \z_j \big) \circ \W \Big]_{h,k} > 0
\end{equation}

\subsubsection{A perceptron.}

\begin{figure}[t]
    \begin{center}
        \includegraphics[height=0.48\textheight]{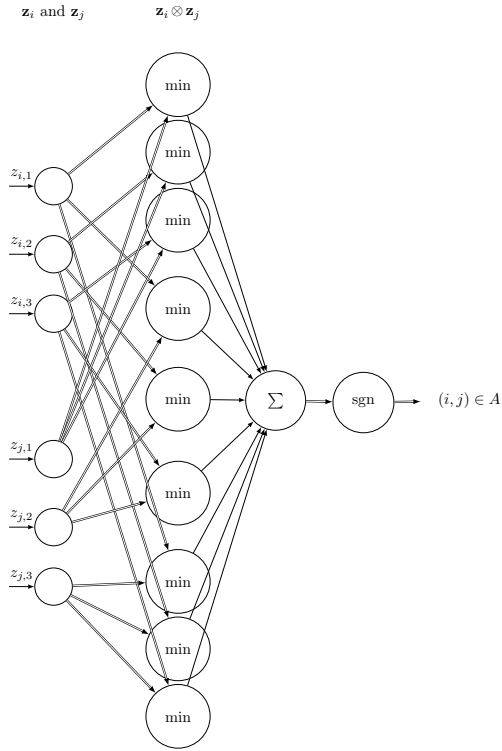}
    \end{center}
    \caption{   \label{fig:llama-nn}
        A neural-network view of the perceptron-like algorithm, for the case of
        $m=3$ features. We indicate fixed weights with double lines, with $\min$
        those nodes activating only if and only if both input nodes are active
        (that is, the $\min$ of their inputs), and with $\sgn$ the sign
        function. The only non-fixed weights (learned by the perceptron update
        rule) are those from the $\z_i \otimes \z_j$ layer to the $\sum$ neuron:
        they correspond to the matrix $\W$ appearing in our model.
    }
\end{figure}

Equation~(\ref{eq:decision-rule}) 
is in fact a special case of the \emph{decision rule} of a
perceptron~\cite{rosenblatt1958perceptron}, the simplest neural network
classifier. The idea here is that by learning how to separate links from
non-links (in fact a form of link prediction), the classifier infers $\W$ as its
internal state.

Let us briefly recall the standard definition: a perceptron is a binary
classifier whose internal state is represented by a vector\footnote{For the
purposes of this paper, we limit ourselves to describing perceptrons with null
bias.} $\textbf{w} \in \mathbb{R}^p$, and it classifies an instance $\textbf{x}
\in \mathbb{R}^p$ as positive if and only if
$\sgn(\textbf{w}\cdot\textbf{x})>0$.

The internal state $\textbf{w}$ is typically initialized at random; then, during
the learning phase, for each $i \in \{ 0,1, \ldots, t-1 \}$:

\begin{enumerate}

    \item the perceptron observes an \emph{example} $\textbf{x}_i \in
    \mathbb{R}^p$;

    \item it emits a \emph{prediction} $\hat{y}_i = \sgn( \textbf{w} \cdot
    \textbf{x}_i ) $;

    \item it receives the \emph{true label} $y_i \in \{-1, 1\}$;

    \item if $y_i \neq \hat{y}_i$, it updates its internal state with
    $\textbf{w} = \textbf{w} + y_i \lambda \textbf{x}_i$, where $\lambda \in (0,
    1]$ is a parameter called \emph{learning rate}.

\end{enumerate}

The key point here is that the decision rule for emitting a prediction can be
cast to be fundamentally the same as in our model. Specifically, if we view the
latent feature-feature matrix $\W$ as a vector of length $m^2$, and we do the
same for $\z_i \otimes \z_j$, then we can see that the decision rule $\sgn(
\textbf{w} \cdot \textbf{x}_i ) = 1$ corresponds to \eqref{eq:decision-rule},
if we set $\W$ as the vector $\textbf{w}$ and $\z_i \otimes \z_j$ as the example
$\textbf{x}$.

Note that in our case an \emph{example} for the perceptron will be a pair of
nodes $(i, j)$, represented not by a vector but by the $m \times m$ matrix $\z_i
\otimes \z_j$: this is a matrix whose element $[\z_i \otimes \z_j]_{h,k}$ is
$1$ if and only if the first node exhibits the feature $h$ and the second
exhibits the feature $k$.
This trick is sometimes called the \emph{outer product kernel}: we are embedding
a pair of vectors of dimension $2m$ into a higher-dimensional representation of
dimension $m^2$.
This $m \times m$-matrix in fact can be alternatively thought of as a vector of size
$m^2$, allowing us to use such vectors as training examples for the perceptron,
where the label is $y=1$ if and only if $(i,j) \in A$, and $y=-1$ otherwise. The
learned vector $\mathbf{w}$ will be, if seen as a matrix, the desired $\W$ appearing in
\eqref{eq:det-model}, as we are going to analyze next.

To recap, the perceptron we are going to use operates like this: given a
$T$ sequence of pairs of nodes (elements of $N \times N$):
\begin{enumerate}

    \item the perceptron observes the next pair $(i,j) \in T$, through their
    binary feature vectors $(\textbf{z}_i,\ \textbf{z}_j)$;

    \item it computes a prediction on whether they form a link, according to
    \eqref{eq:decision-rule}; more precisely, the prediction will be
    $\hat{y}_{i,j} = \sgn(\z_i^T \W \z_j)$

    \item it receives the ground-truth: $y_{i,j} = 1$ if $(i,j) \in A$, and
    $y_{i,j} = -1$ otherwise;

    \item if the prediction was wrong, the updates its internal state by adding
    to $\W$ the quantity
    $y_{i,j} \lambda ( \z_i \otimes \z_j )$.

\end{enumerate}

In doing this, we are using $m^2$ features, in fact a kernel projection of a
space of dimension $2m$ into the larger space of size $m^2$. Similarly, the
weight vector to be learned has size $m^2$. Positive examples are those that
correspond to existing links. We can view this as a shallow, simple neural
network, as depicted in Figure~\ref{fig:llama-nn}.

\paragraph{Interpretation of the error bound.}

One advantage of casting our approach to the perceptron algorithm is that
the latter is a well studied and its performance was analyzed in all details.
In particular, many bounds on its accuracy are known: let us consider
the bounds discussed\footnote{In fact, for the sake of simplicity we are
considering only Euclidean norm and standard hinge loss.} in~\cite[Theorem
12.1]{cesa2006prediction}.
Casting it to our case, some easy manipulations get
the following bound for the number of misclassifications
$M = |\{(i,j) \in T\ s.t.\ \hat{y}_{i,j} \neq y_{i,j}\}| $:
\begin{equation}
\label{eq:bound}
    M \leq \inf_{\textbf{U} \in \mathbb{R}^{m \times m} } \Big(
        H(\textbf{U} ) + \big( R \| \textbf{U} \| \big) ^2 +
            R \| \textbf{U} \| \sqrt{H( \textbf{U} )}
    \Big)
\end{equation}
where $\|-\|$ denotes the Frobenius norm and
\begin{itemize}

     \item $H(\textbf{U} ) = \sum_{(i,j) \in T}{ \max \big( 0, 1 - \z_i^T
     \textbf{U} \z_j \big) } $ is the sum of the so-called \emph{hinge losses}
     and

    \item $R = \max_{(i,j) \in T}{ \| \z_i \otimes \z_j \| }$ is called the
    \emph{radius} of the examples.

\end{itemize}
Let us try to give an interpretation of this bound, by looking at all factors
affecting the number $M$ of errors of the algorithm.
In the following, we want to use the bound above to compute the number of
misclassification which we undergo using \eqref{eq:decision-rule}. For this
purpose, let us set $\textbf{U}=\W$ as in \eqref{eq:decision-rule}.
Suppose also, for the sake of simplicity, that $T=A$ (that is, that we are using all and only the links as examples).
We can define two subsets of $T$:
\begin{eqnarray*}
        E_{\textbf{U}} &=& \big\{ (i,j) \in A\ \big| \ \z_i^T \textbf{U} \z_j
        \leq 0 \big\}\\
        B_{\textbf{U}} &=& \big\{ (i,j) \in A\ \big| \ 0 <\z_i^T \textbf{U} \z_j
        < 1 \big\}.
\end{eqnarray*}
The set $E_{\textbf{U}}$ contains the examples that are \emph{incorrectly
classified} (i.e., those which are not classified as links according to
\eqref{eq:decision-rule}); the set $B_{\textbf{U}}$ contains the examples that
are correctly classified but with a very small margin.
We have that
	\begin{equation}
	\label{eq:hu}
	H(\textbf{U} ) = \sum_{(i,j) \in A}{ \max \big( 0, 1 - \z_i^T
     \textbf{U} \z_j \big) } =\sum_{(i,j) \in E_{\textbf{U}} \cup
     B_{\textbf{U}}} \big( 1 - \z_i^T \textbf{U} \z_j \big) \leq (1+a) \left|E_\textbf{U}\right| + b \left|B_\textbf{U}\right|,
	\end{equation}
for some $a,b>0$ with $b<1$. In other words, the term $H(\textbf{U})$ in the
right-hand-side of (\ref{eq:bound}) is connected with the amount of
misclassifications and borderline-correct classifications: each
misclassification has a cost that is larger than one, whereas borderline-correct
classifications are paid less than one each. In a way, $H(\textbf{U})$ is a
measure of how well our model could fit \emph{in the best case} this particular
feature-rich graph.

One way to reduce the number of borderline-correct classifications would be to
multiply $\textbf{U}$ by a constant larger than one: note that this operation
does \emph{not} change the classification of \eqref{eq:decision-rule}, but at the
same time it increases the cost of misclassifications (the coefficient $a$ of
(\ref{eq:hu})) \emph{and} the norm of $||\textbf{U}||$, that also appears on the
right-hand-side of (\ref{eq:bound}).
The presence of $||\textbf{U}||$ in the bound is explained by the fact that a
model with a large norm is (apart from scaling) more complex: e.g., a very
sparse $\textbf{U}$ (one where only a few pairs of features interact)
will have a very low norm.

The last term appearing in \eqref{eq:bound} is $R^2$, that can be rewritten as

\begin{equation*}
 R^2 = \max_{ (i,j) \in T } \sum_{h} \sum_{k} z_{i,h} z_{j,k} =
        \max_{ (i,j) \in T } | F_i | \cdot | F_j |.
\end{equation*}

In other words, it measures \emph{how many pairs of features} we need to
consider in our set of examples. More precisely, this is the number of possible
pairs among the features of the source and the target of each arc. Of course
$R^2 \leq m^2$: this fact means that the bound is smaller if we need less
features to explain the graph. It is also small if there is little overlap
of features (i.e., if $\max_{ i \in N } | F_i |$ is small).

\medskip

In the case of a feature-rich graph that can be perfectly explained by a latent
feature-feature matrix $\W$ (according to our deterministic model), we have
$H(\W)=0$. In this case, in fact, all the elements of the sum (that is, the
losses suffered by the algorithm) would be null. This can be seen using for
example the inequality given in (\ref{eq:hu}): the set
$\left|E_\textbf{W}\right|$ would be empty, and the same can be said for
$\left|B_\textbf{W}\right|$, possibly scaling $\textbf{W}$ by a constant. In
this special case, the bound simplifies to $M < \big( R \| \textbf{W} \|
\big)^2$. This is the perceptron convergence
theorem~\cite{rosenblatt1961principles}, which in our case tells us that if a
perfect $\W$ exists, the algorithm will converge to it.

\subsubsection{A passive-aggressive algorithm}
\label{sec:pa-llama}

\paragraph{Online learning.}

In general, what we did was to recast our goal in the framework of online binary
classification. Binary classification, in fact, is a well-known problem in
supervised machine learning; \emph{online} classification simplifies this
problem by assuming that examples are presented in a sequential fashion and that
the classifier operates by repeating the following cycle:

\begin{enumerate}
    \item it observes an example;
    \item it tries to predict its label;
    \item it receives the true label;
    \item it updates its internal state consequentially, and moves on to the
    next example.
\end{enumerate}

An online learning algorithm, generally, needs a constant amount of memory with
respect to the number of examples, which allows one to employ online algorithms
in situations where a very large set of voluminous input data is available. A
survey is available in~\cite{cesa2004generalization}.

A well-known type of online learning algorithms are the so-called
perceptron-like algorithms. They all share the same traits of the perceptron:
each example must be a vector $\mathbf{x}_{i}\in\mathbb{R}^{p}$; the internal
state of the classifier is also represented by a vector
$\mathbf{w}\in\mathbb{R}^{p}$; the predicted label is $y_i=sign(\mathbf{w} \cdot
\mathbf{x}_{i} )$. The algorithms differ on how $\mathbf{w}$ is built. However,
since their decision rule is always the same, they all lead back to the decision
rule of our model~\eqref{eq:decision-rule}. This observation allow us to employ
\emph{any} perceptron-like algorithm for our purposes.

Perceptron-like algorithms (for example, \textsc{ALMA}~\cite{GenALMA} and
Passive-Aggressive~\cite{CDKSSOPAA}) are usually simple to implement, provide
tight theoretical bounds, and have been proved to be fast and accurate in
practice.

\paragraph{A Passive-Aggressive algorithm.}

Among the existing per\-cep\-tron-like online classification frameworks, we will
heavily employ the well-known Passive-Aggressive classifier, characterized by
being extremely fast, simple to implement, and shown by many
experiments~\cite{CARVALHO2006SINGLE, MRZZAC} to perform well on real data.

\begin{algorithm}[t]
		\textsc{Input}: \\
		\-\hspace{0.4cm} The graph $G=(N, A)$, with $A \subseteq N \times N$ \\
		\-\hspace{0.4cm} Features $F_i \subseteq F $ for each node $i \in N$
		\\
		\-\hspace{0.4cm} A parameter $\kappa>0$ \\
		\textsc{Output}: \\
		\-\hspace{0.4cm} The feature-feature latent matrix $\W$ \\
		\begin{enumerate}
			\item $\W \leftarrow \mathbf{0}$
            \item Let $(i_1,j_1),\dots,(i_T,j_T)$ be a sequence of
                  elements of $N \times N$.
			\item For $t=1,\dots,T$

			\begin{enumerate}
			    \item $\rho \leftarrow 1/(|F_{i_t}| \cdot |F_{j_t}|) $
				\item $\mu \leftarrow \sum_{h \in F_{i_t} }{ \sum_{k \in F_{j_t}} {
				W_{h, k} }}$
				\item \textbf{If} $(i_t,j_t) \in A$ \\
						\-\ $\quad \delta \leftarrow \min(\kappa, \max(0, \rho (1-\mu) ) )$ \\
					  \textbf{else} \\
						\-\ $\quad \delta \leftarrow - \min(\kappa, \max(0, \rho (1+\mu) ) )$ \\
				\item For each $h \in F_{i_t}$,  $k \in F_{j_t}$: \\
						\-\ $\quad W_{h,k} \leftarrow W_{h,k} + \delta $
			\end{enumerate}
		\end{enumerate}
\caption{   \label{alg:our-pa-alg}
    \\ \textsc{Llama}, the passive-aggressive algorithm to build the latent
    feature-feature matrix $\W$.
}
\end{algorithm}

Le us now describe the well-known Passive-Aggressive algorithm~\cite{CDKSSOPAA},
while showing how to cast this algorithm for our case. To do this let us
consider a sequence of pairs of nodes
\[
	(i_1,j_1),\dots,(i_T,j_T) \in N \times N
\]
(to be defined later). Define a sequence of matrices $\W^0,\dots,\W^T$ and of
slack variables $\xi_1,\dots,\xi_T \geq 0$ as follows:
\begin{itemize}
  \item $\W^0=\textbf{0}$
  \item $\W^{t+1}$ is a matrix minimizing $\|\W^{t+1}-\W^t\|+\kappa\xi_{t+1}$
  subject to the constraint that
  \begin{equation}
  \label{eqn:ineq}
      y_{i_t,j_t} \cdot \sum_{h \in F_{i_t}} \sum_{k \in F_{j_t}}
      W^{t+1}_{h,k}\geq 1 - \xi_{t+1},
  \end{equation}
where, as before
  \[
  	y_{i_t,j_t}=\begin{cases}
  					-1	& \text{if $(i,j) \not\in A$}\\
  					1	& \text{if $(i,j) \in A$}
  				\end{cases},
  \]
$\|-\|$ denotes again the Frobenius norm and $\kappa$ is an optimization
parameter determining the amount of aggressiveness.
\end{itemize}

The intuition behind the above-described optimization problem, as discussed
in~\cite{CDKSSOPAA}, is the following:

\begin{itemize}

  \item the left-hand-side of the inequality (\ref{eqn:ineq}) is positive if and
  only if $\W^{t+1}$ correctly predicts the presence/absence of the link
  $(i_t,j_t)$; its absolute value can be thought of as the confidence of the prediction;

  \item we would like the confidence to be at least 1, but allow for some error
  (embodied in the slack variable $\xi_{t+1}$);

  \item the cost function of the optimization problem tries to keep as much
  memory of the previous optimization steps as possible (minimizing the
  difference with the previous iterate), and at the same time to minimize the
  error contained in the slack variable.

\end{itemize}

By merging the Passive-Aggressive solution to this problem with our
aforementioned framework, we obtain the algorithm described in
Algorithm~\ref{alg:our-pa-alg}. We will refer to this algorithm
as~\textsc{Llama}:~\emph{Learning LAtent MAtrix}.


\paragraph{Normalization.}

For perceptron-like algorithms, normalizing example vectors (in our case, the
matrix $\z_i \otimes \z_j $) often gives better results in
practice~\cite{cristianini2000introduction}. This is equivalent to using the
$\ell^2$-row-normalized version of our model, as discussed in
\sec{par:row-normalization} (setting $p=2$). The assumption behind that model is
in fact that nodes with fewer features provide a stronger signal for the small
set of features they have; nodes with many features bear less information about those feature.

It is immediate to see that Algorithm~\ref{alg:our-pa-alg} can be adapted to use
the $\ell^2$-row-normalization by changing step (c) to:
\begin{quotation}
\begin{enumerate}[(c)]
    \item \textbf{If} $(i_t,j_t) \in A$: \\
            \-\ $\quad \delta \leftarrow \sqrt{\rho} \min(\kappa,
            \max(0, 1-\sqrt{\rho}\mu ) )$ \\
          \textbf{else}:  \hspace*{\fill}  \eqnum\label{eq:normalized-llama} \\
            \-\ $\quad \delta \leftarrow - \sqrt{\rho} \min(\kappa,
            \max(0, 1+\sqrt{\rho}\mu ) )$ \\
\end{enumerate}
\end{quotation}
Similar adaptations would allow one to implement \emph{any} row normalization.

\paragraph{Sequence of pairs.} Finally, let us discuss how to build the sequence of
examples. We want $\W$ to be built through a single-pass online learning
process, where we have all positive examples at our disposal (and they are in
fact all included in the training sequence), but where negative examples cannot
be all included, because they are too many and they would produce overfitting.

Both the Passive-Aggressive construction described above and the Perceptron
algorithm depend crucially on the sequence of positive and negative examples
$(i_1,j_1),\dots,(i_T,j_T)$ that is taken as input. In particular, as discussed
in~\cite{JAPKOWICZ2002CLASSIMBALANCE}, it is critical that the number of
negative and positive examples in the sequence is balanced. Taking this
suggestion into account -- and also considering~\cite{yang2015evaluating}
suggestions about uniform sampling -- we build the sequence as follows: we  draw uniformly
at random $|A|$ node pairs $(i,j)$ s.t. $(i,j) \notin
A$; then, nodes are enumerated (in arbitrary order), and for each node $i \in
N$, all arcs of the form $(i,\bullet)\in A$ are added to the sequence, followed
by all non-links node pairs of the form $(i,\bullet)$. Of course, in the end the
sequence contain $T=2\cdot |A|$ node pairs -- that is, $|A|$ links along with
$|A|$ non-links.

Obviously, there are other possible ways to define the sequence of examples and
to select the subset of negative examples. However, we chose to adopt this
technique (single pass on a balanced random sub-sample of pairs) in order to
define and test our methodology with a single, natural and computationally
efficient approach. However, when experimenting with real data in
Section~\ref{sec:real-data}, we will also test whether the ordering of nodes
affects the results, by comparing natural (i.e. chronological) and random order.

\paragraph{Error bound for Passive-Aggressive.}
The analysis of the error bound for misclassifications of the perceptron
(\ref{eq:bound}) can be made more precise for the case of the
Passive-Aggressive algorithm: using Theorem 4 of~\cite{CDKSSOPAA},
the bound becomes:
\begin{equation}
\label{eq:boundpa}
    M \leq \inf_{\textbf{U} \in \mathbb{R}^{m \times m} } \max\left(R^2,
    1/\kappa\right) \Big( 2 \kappa H(\textbf{U} ) + \| \textbf{U} \| ^2
    \Big).
\end{equation}
If $\kappa=1/R^2$, the bound reduces to
\[
    M \leq \inf_{\textbf{U} \in \mathbb{R}^{m \times m} } 2
    H(\textbf{U} ) + \left( R \| \textbf{U} \| \right) ^2,
\]
and our discussion of (\ref{eq:bound}) is essentially confirmed. We encounter
\mbox{$R^2=\max_{ (i,j) \in T } | F_i | \cdot | F_j |$,} that is the maximum
number of pairs of features we observe at the same time; $H(\textbf{U} )$, the
total loss of the ``best'' (in terms of the infimum in the equation) possible
feature-feature matrix; and $\| \textbf{U} \|$, the norm of such a matrix, which
is fundamentally a measure of its complexity. Also for Passive-Aggressive, these
factors define the performance of the algorithm on a specific instance of
feature-rich graph.

A truly on-line approach with
unnormalized samples will require a constant $\kappa$  (in our experiments we
set $1.5$), which yields
\[
    M \leq \inf_{\textbf{U} \in \mathbb{R}^{m \times m} }
    c R^2 H(\textbf{U} ) + \left( R \| \textbf{U} \| \right) ^2,
\]
for some constant $c$.

\section{Experiments on synthetic data}
\label{sec:llama-simulations}

In this section, we will test how the methods described in this paper perform on
synthetic graphs generated within our framework using the techniques described
in previous work~\cite{bolcrimon}; in the next section we will see how they
behave on real-world data.

We are in fact building upon previous methods~\cite{bolcrimon} to generate a
realistic node-feature association $\Z$ that, when used as input to the model of
\eqref{eq:basic-model}, is able to synthesize feature-rich networks with the
same traits  (e.g., distance distribution, degree distribution, fraction of
reachable pairs, etc) as typical real complex networks. In particular,
in~\cite{bolcrimon} we discuss how to generate a synthetic feature-rich graph
with the same properties as a given real one. These experiments allow us to
employ graphs generated through this approach as a test bed for the algorithms
presented in the Section~\ref{chap:inferring-w}.

\subsection{Experimental setup}
\label{sec:simulated-networks}

\begin{table}
    \begin{center}

            \textbf{Avg. features per node}  \\

            \begin{tabular}{ llll }
                \toprule
                        & $S$    & $\chi$ & $\exp$ \\ \midrule
               $\bern$  & $5.84 \pm 1.63$ & $5.17 \pm 1.63$ & $5.22 \pm 1.51$ \\
               $\Gauss$ & $5.76 \pm 1.43$ & $5.30 \pm 1.56$ & $5.51 \pm 1.31$ \\
               \bottomrule
            \end{tabular}

            \vspace{5mm}

            \textbf{Avg. degree}             \\

            \begin{tabular}{ llll }
                \toprule
                        & $S$     & $\chi$  & $\exp$        \\ \midrule
               $\bern$  & $109.4 \pm 325$ & $163.2 \pm 329$ & $15.6 \pm 217$ \\
               $\Gauss$ & $ 10.9 \pm 145$ & $11.8  \pm 138$ & $26.3 \pm 299$ \\
               \bottomrule
            \end{tabular}

            \vspace{5mm}

            \textbf{Mean harmonic distance}  \\

            \begin{tabular}{ llll }
                \toprule
                        & $S$    & $\chi$ & $\exp$           \\ \midrule
               $\bern$  & $2.16 \pm 92$ & $2.43 \pm 1\,339$ & $2.02 \pm 3\,290$ \\
               $\Gauss$ & $2.31 \pm 3\,034$ & $11.0 \pm 2\,472$ & $2.01 \pm 1\,606$ \\
               \bottomrule
            \end{tabular} \\

        \caption{   \label{tab:simulation-prop} Properties of the synthetic
        feature-rich graphs. The $6$ generated graph families are indicated
        according to the $\phi$ function used ($S$ is the sigmoid, $\chi$ is the
        step function, and $\exp$ is the exponential) and to the distribution of
        the values of $\W$ (Bernoullian or normal). The listed properties
        represents the median, inside each graph family, of: the average number
        of features per node, the average degree and and the mean harmonic
        distance. }

\end{center}
\end{table}

To generate each network, we first produced its node-feature association $\Z$
with the Indian Buffet Model method~\cite{bolcrimon}, using the same parameter
values adopted in previous work: $\alpha=3$, $\beta=0.5$, $c=0$. Then, we fed
these matrices $\Z$ to our model equation \eqref{eq:basic-model} to generate a
number of graphs. For the graph model, we employed the following parameters:

\begin{itemize}

    \item We used $n=10\,000$ nodes.

    \item We applied three different types of activation function
    $\phi$, to compare their results:
    \begin{enumerate}

        \item The classic sigmoid function $S(x)=\big( e^{K(\theta-x)}+1 \big)
        ^{-1}$, cited in \sec{sec:model-parameters} as well as
        in~\cite{bolcrimon} as the standard approach; we set $\theta=0$ and
        $K=5$.
        Please note that this function does \emph{not} respect the assumptions
        for which we derived \textsc{Llama}, nor those of \textsc{Naive}.

        \item The step function $\chi_{(0,\infty)}$, characterizing the model
        behind \textsc{Llama}.

        \item The $\exp$ function, which characterizes the model behind
        \textsc{Naive}.

    \end{enumerate}

    \item The latent matrix $\W$ was generated assuming that its entries are
    i.i.d., with the following two value distributions:

    \begin{enumerate}

        \item A generalized Bernoulli distribution $W_{h,k} \sim \mathcal{B}(p)$
        that assumes the value $10$ with probability $p=\frac{10}{m}$  and $-1$
        with probability $1-p$. This choice was determined through experiments,
        with the purpose of obtaining graphs with a realistic density independently from
        the number of features $m$.

        \item A normal distribution $W_{h,k}\ \sim\ \mathcal{N}(\mu,\,\sigma)$
        with mean and variance identical to the previous Bernoulli
        distribution.

        \item We had to slightly modify these distributions for the case
        $\phi=\exp$, in order to obtain realistic graphs
    	also in that case: in particular, when $\phi=\exp$ we used a Bernoulli
    	distribution with value $1$ with probability $p=\frac{1}{m}$ and $-1$ with
    	probability $1-p$, and a normal distribution that had the same
    	mean and variance as the just-described Bernoulli distribution. In the
    	following, when we say that $\phi=\exp$ we imply that we used one of these
    	two modified distributions to generate $\W$.
    \end{enumerate}

\end{itemize}

With these three choices for $\phi$ and two choices for the generation of $\W$,
we obtained six different families of feature-rich graphs. For each graph
family, we generated $100$ different graphs. The properties of these networks
are summed up in Table~\ref{tab:simulation-prop}. They represent a wide range of
realistic traits we could actually observe in complex networks.

\subsection{Evaluation}
\label{sec:evaluating-lp}

First of all, even if the aim of both \textsc{Llama} and \textsc{Naive} is to
reconstruct the matrix $\W$, we are not interested in the actual values
of the elements of $\W$. Our goal is to find a feature-feature matrix for which our
model works: it is not important if the values are scaled up or shifted as long
as the predictions of our model for the links remain correct.

For this reason, we will measure directly how accurate our methods are in terms
of predicting if a node pair $(i,j)$ forms a link, given their features. To keep
this evaluation meaningful, our algorithms will not be allowed to see the whole
graph: we will use the standard approach of $10$-fold cross-validation; i.e., we
divide the set of nodes $N$ into ten subsets (folds) of equal size, and
we used nine folds to train the algorithm and the tenth remaining fold to test
the results (for each possible choice of the latter).

Our evaluation closely resembles the approach followed for \emph{link
prediction}. There are of course some differences: first of all, we are using an
external source of information (the node features) that is not available to
link-prediction methods; second, our aim is to evaluate our model and our
algorithms to find $\W$ \emph{through} link prediction. That is, we are not
interested in finding the best existing link predictor, but in measuring if our
algorithms can correctly fit our model on a specific instance of feature-rich
graph $(G,\Z)$. However, we followed the evaluation guidelines for link
prediction recently stated by Yang~\emph{et al.}~\cite{yang2015evaluating}.

\begin{itemize}

    \item We evaluated how accurate our algorithms are in prediction by showing
    precision/recall curves: Yang~\emph{et al.}~\cite{yang2015evaluating}, in
    fact, observe that other alternatives, such as the ROC curve, are heavily
    biased after undersampling negative examples and can yield misleading
    results; since tied scores do affect results (especially for
    \textsc{Naive}), we employed the techniques described
    in~\cite{mcsherry2008computing} to compute precision and recall values for
    tied scores.

    \item Using precision/recall curves allow us to avoid using a fixed
    threshold between ``link''/``not link''; it is important, in fact, to
    evaluate the \emph{scores} themselves; on the contrary, by choosing a
    threshold $\theta$ and then converting each score $x$ to a binary event $x >
    \theta$ would make the comparison unfair; we instead used directly the score
    computed by our model (the argument of $\phi$ in \eqref{eq:basic-model})
    since the larger this score, the more probable that link should be.

    \item We used the same test set for all the tested algorithms.

    \item Although in our case it was necessary to undersample negatives (the
    total number of node pairs would be unmanageable), we took care of sampling
    uniformly the edges missing from the test network: we draw node pairs
    $(i,j)$ such that $(i,j) \notin A$ uniformly from the set $N \times N$,
    until we had a number of non-arcs equal to the number of arcs.

\end{itemize}

Since our methods are not influenced by the distances of the pairs of nodes
involved (contrarily to standard link prediction approaches), we
avoided to gather our results by geodesic distance.

With the above considerations in mind, we proceeded to evaluate our approach
using precision-recall curves. For each of the networks and for each fold, we
gave the training graph as input to the algorithm (\textsc{Llama} or
\textsc{Naive}) and obtained an estimated matrix $\W$. This matrix is defined
by\footnote{In the case $(N_h \times N_k) \cap A = \emptyset$, the Naive
approach as described by \eqref{eq:naive} would set $W_{h,k}=\log 0$. We tried
two alternative strategies to solve this issue: (i) setting $W_{h,k}$ equal to a
large negative number for those pairs (\emph{de facto} putting a lower bound to
$W_{h,k}$); (ii) employing an add-one smoothing~\cite{russellnorvig}, i.e.,
using $\log(x+1)$ in place of $\log(x)$. The experimental results are
essentially the same in the two cases. The figures presented in this section are
the ones obtained by (i).} \eqref{eq:naive} for \textsc{Naive} and
by\footnote{We tried also the normalized version of \textsc{Llama} expressed in
\eqref{eq:normalized-llama}, for different values of $p$, leaving the model
unchanged. Again, the experimental results obtained are the same on our dataset,
so we are here presenting the values obtained by the unnormalized version of the
algorithm.} Algorithm~\ref{alg:our-pa-alg} for \textsc{Llama}. Each method then
assigned its score (i.e., the argument of $\phi$ in \eqref{eq:basic-model}) to
each node pair in the test set, according to our model.

\newcommand{\simulationresult}[2]{
	\begin{center}
		\begin{tabular}{cc}
			\includegraphics[width=0.5\columnwidth]{img/#1-10k-1-prec-rec-NaiveEstimator}
			&
			\includegraphics[width=0.5\columnwidth]{img/#1-10k-1-prec-rec-LlamaEstimator} \\
            \textsc{Naive} & \textsc{Llama}
		\end{tabular}
		\caption{\label{fig:#1} Precision-recall curves
	in the network $#2$. Different colors represent different folds used in cross-validation. }
    \end{center}
}

\subsection{Training time}
\label{sec:evaluating-lp}

\begin{table}[t]
    \begin{center}
    \begin{tabular}{ lll }
        \toprule
          & AUPR & Time (s) \\
         \midrule
         \textsc{Naive}
            & $0.824 \pm 0.028$ 
            & $0.034 \pm 0.034$ 
\\         \textsc{Llama}
            & $0.893 \pm 0.020$ 
            & $0.097 \pm 0.097$ 
\\         \textsc{SVM}
            & $0.915 \pm 0.014$ 
            & $6439.303 \pm 6439.303$ 
\\
         \bottomrule
    \end{tabular}

    \caption{   \label{tab:simulation-times} Area under the precision-recall
    curve (on average across 10 folds and 4 experiments) and the required
    training time in seconds. For each value we report the mean and the
    standard deviation. }

\end{center}
\end{table}

Before discussing the results, let us present a measure of the training times of
the algorithms we propose, in comparison with SVM, a baseline previously
employed in the literature for feature-rich graphs~\cite{doppa2010learning}. For
this algorithm, we are using an efficient implementation (the one from
WEKA~\cite{hall2009weka}), written in the same language as our own algorithms,
and using therefore the same methods for I/O. We employed a linear kernel (the
fastest) for the SVM.

The results we show are about a single graph family of the ones discussed above
(specifically, the case where $W_{h,k} \sim \mathcal{N}(\mu,\,\sigma)$ and the
sigmoid function $S(x)$ is used as an activation function). These are the most
common cases treated in the literature. Also, we needed to set a lower number of
nodes $n=1000$ in order for the SVM to terminate.

Our results (Table~\ref{tab:simulation-times}) show a training time for the SVM
that is four orders of magnitude longer than \textsc{Naive} or \textsc{Llama},
i.e.,
taking on the scale of hours for graphs of thousands of nodes. These results are
consistent with the previous literature. Perceptron-like algorithms are known to
be much less computationally expensive than traditional SVMs
\cite{sculley2007online}. However, despite them to be unusable at the scale we
want to operate (tens millions of nodes), it is worth noting that their
performance is (slightly) better than \textsc{Llama} in this particular case.

\subsection{Results}

We report detailed performance results for \textsc{Naive} and \textsc{Llama} in
Table~\ref{tab:simulation-aupr}. There, we show the average AUPR (Area Under
Precision-Recall curve) obtained across all the graphs inside each graph family
considered. To compute the AUPR we used the technique described by Davis and
Goadrich~\cite{davis2006relationship}.

Following the previous suggestions~\cite{yang2015evaluating}, we use this area
as an overall measure of the goodness of our approach. We can see how the
results of \textsc{Llama} are on average above $95\%$ for both the step function
and the sigmoid activation function. The $\exp$ case, in fact, is the one where
\textsc{Naive} works better -- as it was expected from the theory. In the normal
case, the performance of \textsc{Llama} is still good; a Bernoulli distribution
with an exponential activation function is instead the only case when
\textsc{Llama} performance is inadmissible. As we shall see in
Section~\ref{sec:real-data}, though, the $\exp$ case does not correspond to a
realistic setting.

\begin{table}[tb]
    \begin{center}
    \begin{tabular}{ lllllllll }
        \toprule
          &  $S,\bern$ & $S,\Gauss$ & $\chi,\bern$ & $\chi,\Gauss$ & $\exp,\bern$ & $\exp,\Gauss$ \\
         \midrule
            \textsc{Naive}
            & $.843 \pm .060$ 
            & $.951 \pm .148$ 
            & $.599 \pm .288$ 
            & $.798 \pm .258$ 
            & $.931 \pm .232$ 
            & $.972 \pm .084$ 
            \\	\textsc{Llama}
            & $.974 \pm .016$ 
            & $.951 \pm .151$ 
            & $.973 \pm .018$ 
            & $.967 \pm .117$ 
            & $.529 \pm .279$ 
            & $.880 \pm .155$ 
            \\
           \bottomrule
        \end{tabular}

    \caption{   \label{tab:simulation-aupr} Area under the precision-recall
    curve of \textsc{Naive} and of \textsc{Llama}. For each of the considered
    graph families, we report the mean and the standard deviation across all the
    graphs. }

\end{center}
\end{table}

Let us discuss in details the results obtained, gathering them by the activation
function employed to generate the graph. To be able to grasp what happens across
different folds in a single graph, and to avoid overcrowding the plots, we will
report the precision-recall plots for a single graph inside each family.

\paragraph{Step activation (Figure~\ref{fig:step-bernoulli}
and~\ref{fig:step-normal}).}

\begin{figure}[th]
    \simulationresult{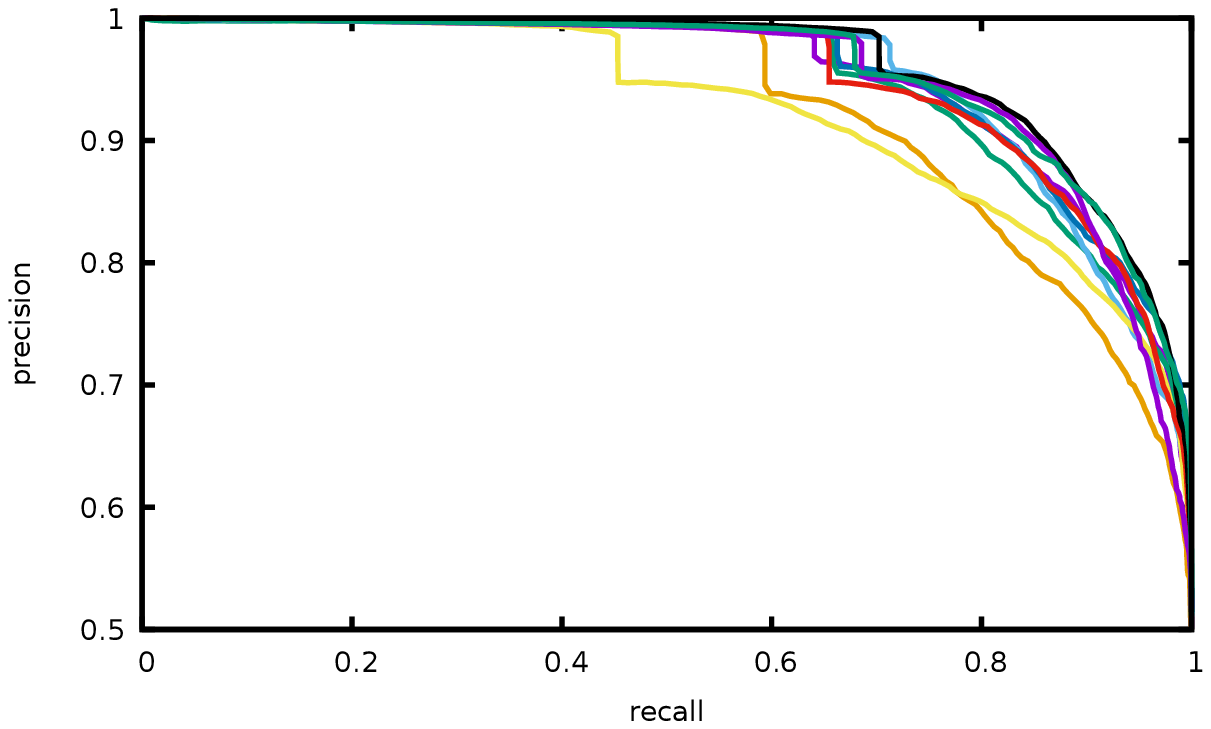}{\chi,\mathcal{B}}
    \simulationresult{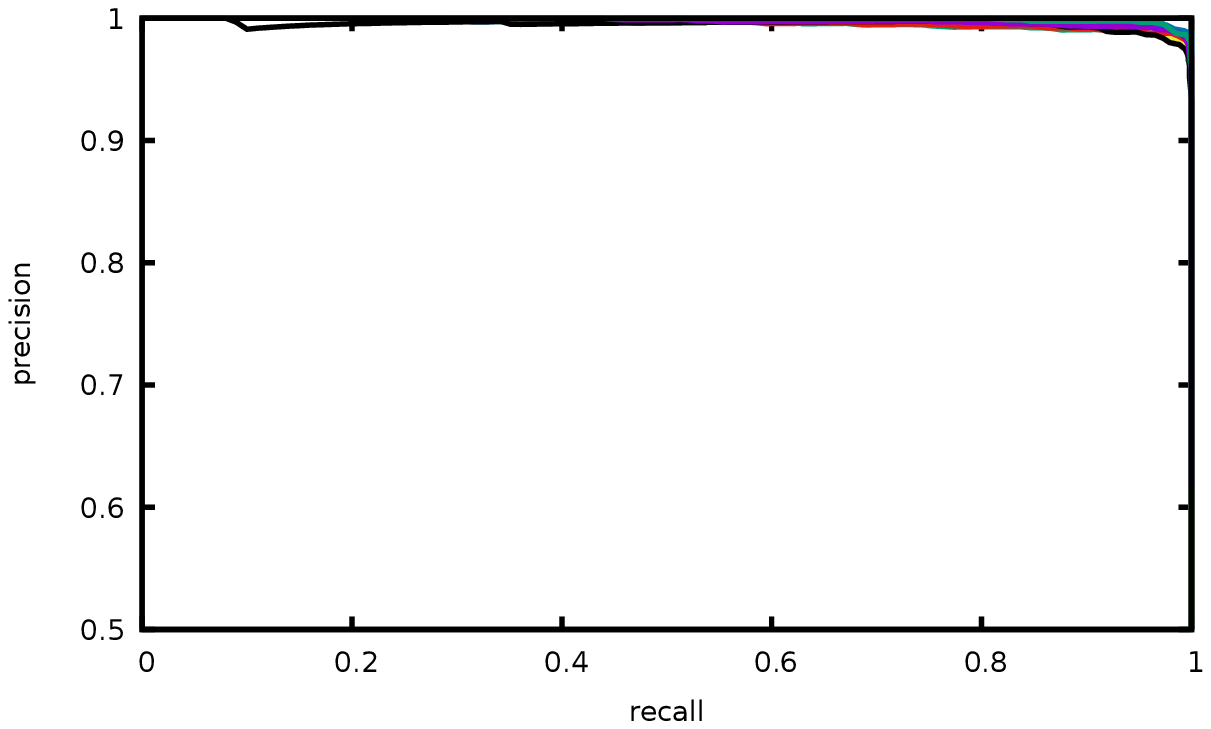}{\chi,\mathcal{N}}
\end{figure}

Let us first consider the case of the networks generated with a step activation
function. Note that by using $\chi_{(0,\infty)}$ as the activation function we
are making our model deterministic --- a pair forms a link if and only if its
score is positive. Furthermore, this is precisely the activation function for
which we have formal guarantees on the \textsc{Llama} performances. In
fact, its results are remarkably good, as testified by an area under curve
beyond $96\%$ in both the Bernoullian and the Gaussian case.

\textsc{Naive} is able to take advantage of this clean activation function only
with a normal distribution on the values of $\W$ (where its performance is
around $80\%$); in the bernoullian case, it degrades toward a random classifier.

\paragraph{Exponential activation (Figure~\ref{fig:exp-bernoulli}
and~\ref{fig:exp-normal}).}

\begin{figure}[t]
    \simulationresult{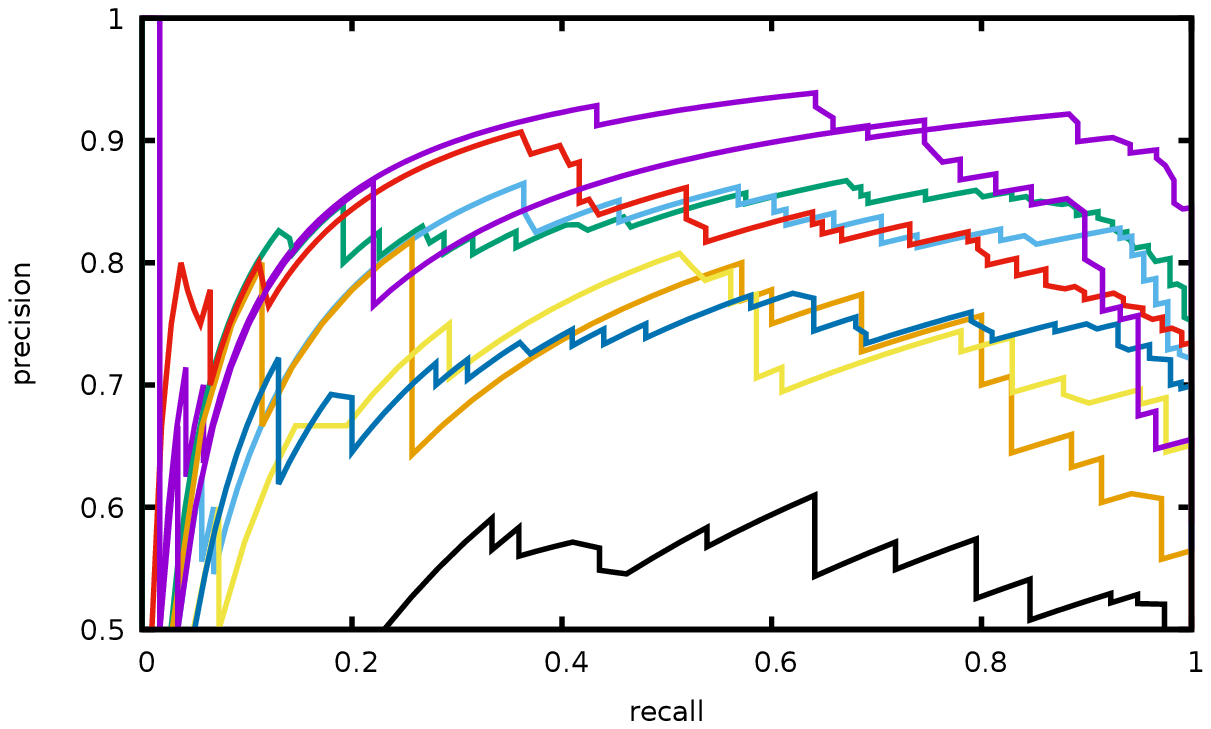}{\exp,\mathcal{B}}
    \simulationresult{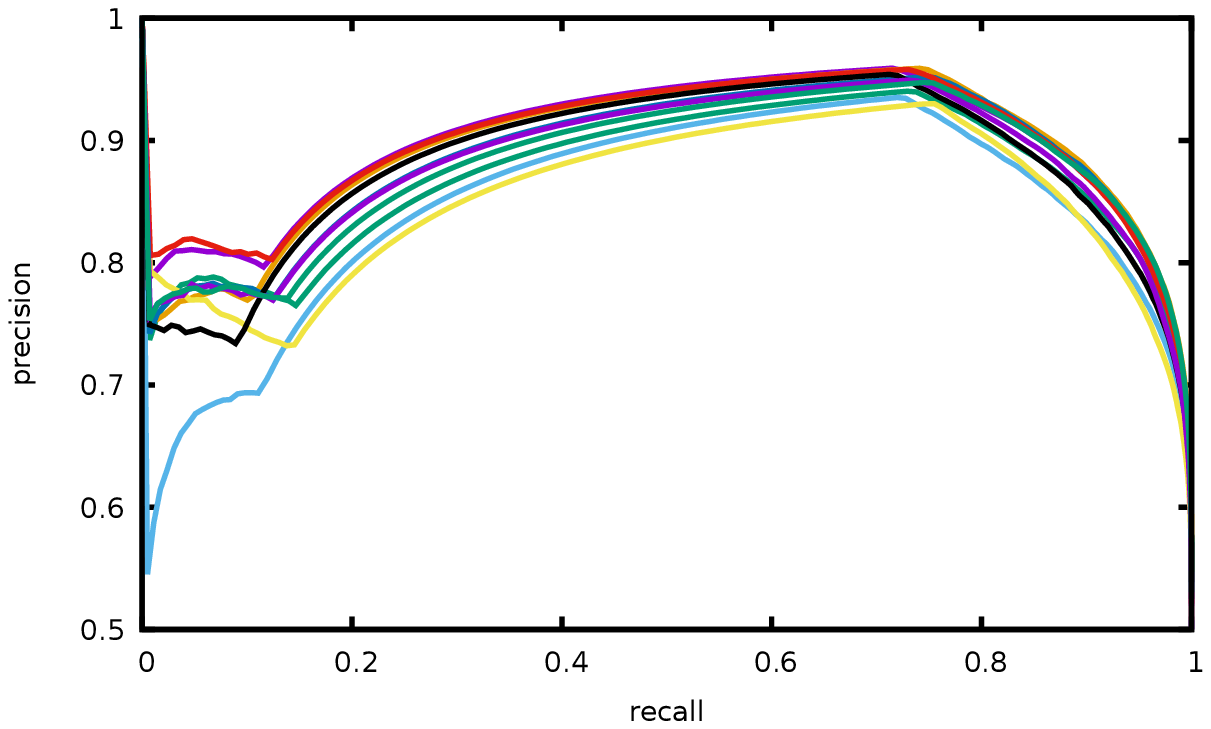}{\exp,\mathcal{N}}
\end{figure}

Let us now look at the exponential activation function, for which we have
formally derived \textsc{Naive}. The results obtained by \textsc{Naive} are in
fact very good at all recall levels.

\textsc{Llama}, on the other hand, obtains its worst performance on this
simulation, due to the fact that the exponential function is mostly dissimilar
from \textsc{Llama}'s natural one (the step function). In the bernoullian case
its performance is chaotic, and depends very much on the training set; instead,
in the normally-distributed case, the area under the precision-recall curve is
definitely better, around $80\%$ on average.

\paragraph{Sigmoid activation (Figure~\ref{fig:sigmoid-bernoulli}
and~\ref{fig:sigmoid-normal}).}

\begin{figure}[t]
    \simulationresult{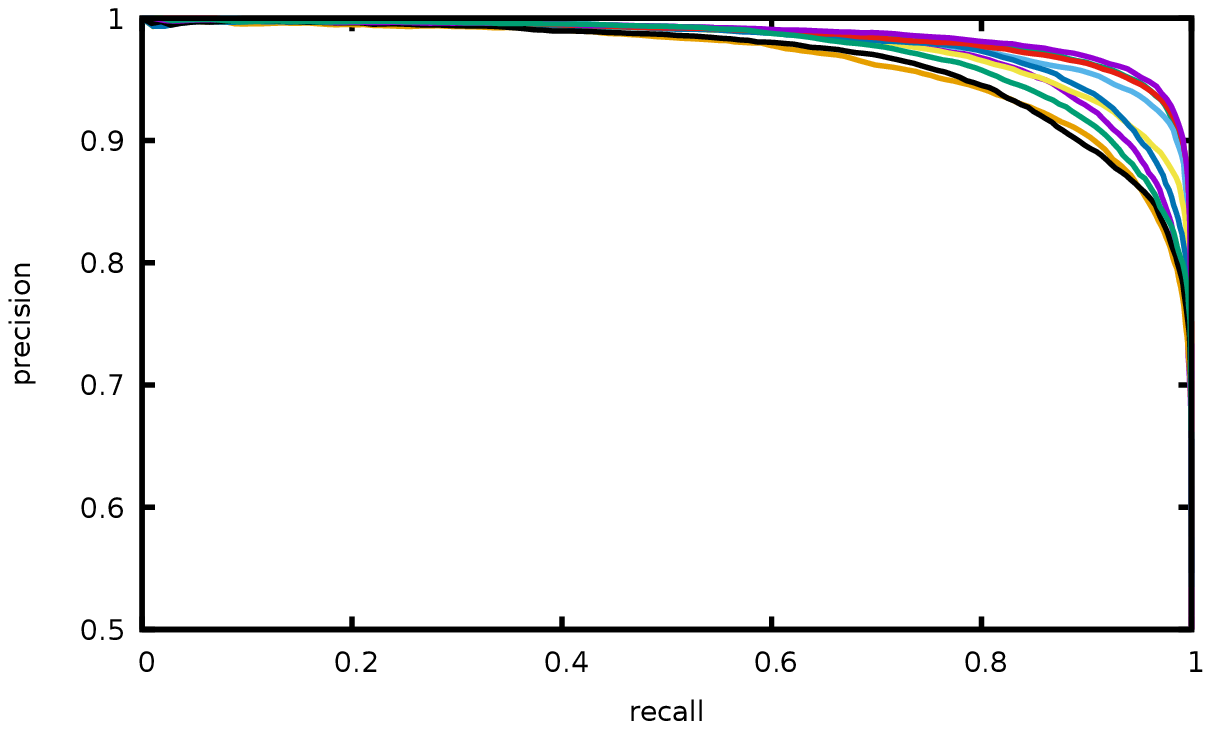}{S,\mathcal{B}}
    \simulationresult{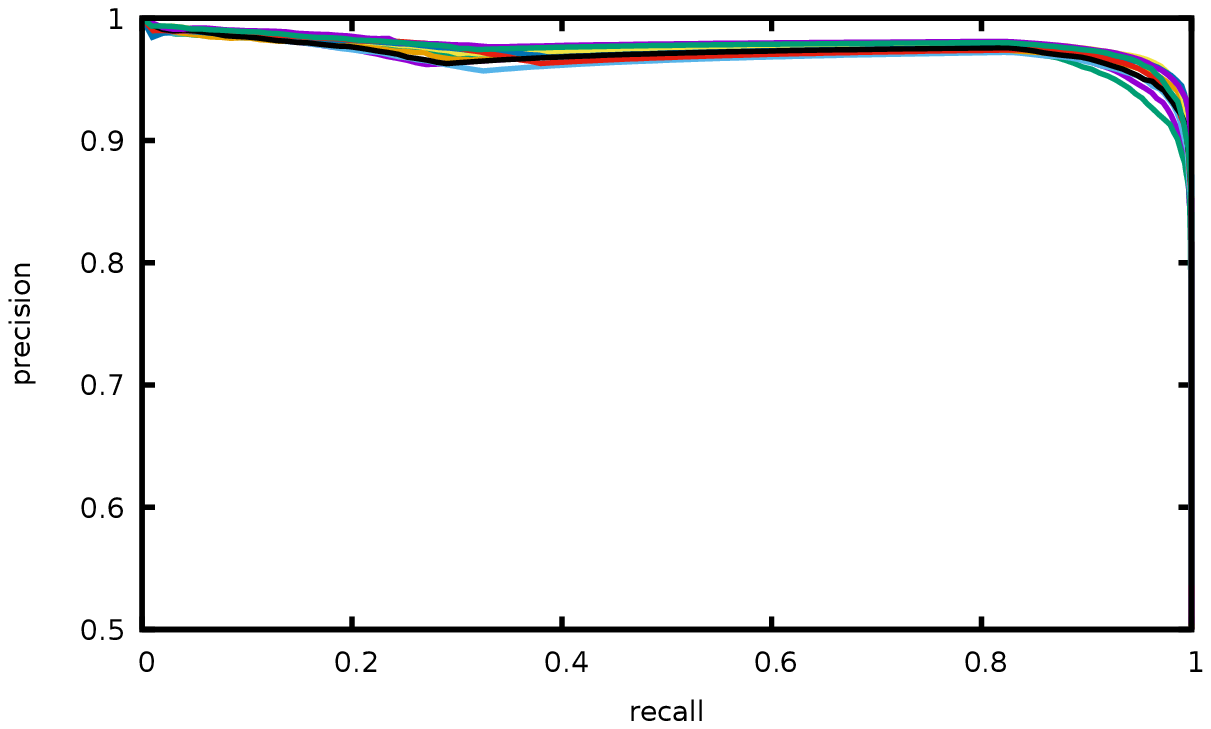}{S,\mathcal{N}}
\end{figure}

Finally, let us look at the results obtained when the activation function is a
sigmoid \eqref{eq:sigmoid} with $K=5$. We emphasize that this activation
function is one for which we have no theoretical guarantees, neither for
\textsc{Llama} (which assumes a step function) nor for \textsc{Naive} (which
assumes an exponential); also, it is the function of choice in previous
literature (e.g.~\cite{MGJ2009}).

We report in Figure~\ref{fig:sigmoid-bernoulli} the precision-recall curves for
the case of the Bernoulli distribution and in Figure~\ref{fig:sigmoid-normal}
the precision-recall curves for the case of the normal distribution. We can see
how \textsc{Naive} performances display a high variance and are way behind the
\textsc{Llama} performances, especially in the Bernoulli-distributed case.
\textsc{Llama} performances in fact are almost as good as in its natural step
function case, with an area under curve consistently beyond $95\%$.

The unambiguous prevalence of \textsc{Llama} in this ``natural'' case could
explain the results we are showing in the next section.

\section{Experiments on real data}
\label{sec:real-data}

In this section, we will focus on (1) how our algorithms behave on real-world
feature-rich networks and (2) how our framework can be used to evaluate
the relationship between a network and a particular set of features for its
nodes. In particular, we will consider the fitness of our model as a measure of
how much a certain set of features can explain the links in such a graph.

\paragraph{Explainability.}

Given a graph $G=(N,A)$ and a particular set of features $\hat{F}$ that can be
associated to its nodes (with $\hat{Z} \subseteq N \times \hat{F}$), we can
define the \emph{explainability} of $\hat{F}$ for $G$ to be the area under the
precision-recall curve obtained by the scores provided by our model; with
``score'' we mean, as before, the argument of $\phi$ in \eqref{eq:basic-model},
where the matrix $\W$ is the one found by Algorithm~\ref{alg:our-pa-alg} when it is given $G$ and $\hat{Z}$ as
input. We again use the AUPR (Area Under Precision-Recall curve) as a measure of
fitness, as we did in \sec{sec:evaluating-lp}.


\subsection{Experimental setup}

We are going to consider a scientific network recently released by Microsoft
Research, and known as the Microsoft Academic Graph~\cite{sinha2015overview}. It
represents a very large (tens of millions), heterogeneous corpus of scientific
works; each scientific work has some metadata associated with it.

We will consider the citation network formed by these papers: this is a directed
graph whose nodes are the papers, and with an arc $(i,j) \in A$ if and only if
paper $i$ contains a citation to paper $j$. As for the features, we will consider the following alternative sets
of node features:

\begin{itemize}
    \item authors' \emph{affiliations}: for each paper, all the institutions
    that each author of the paper claims to be associated to.
    ``University of Milan'' and ``Google'' are examples of affiliations.
    \item the set of \emph{fields of study}: the field of study associated by
    the dataset curators~\cite{sinha2015overview} to the keywords of the paper.
    ``Complex network'' and ``Vertebrate paleontology'' are examples of fields
    of study.
\end{itemize}

These features fully respect all the assumptions we made: they are attributes of
the nodes, they are binary (a node can have a feature or not, without any middle
ground), they are possibly overlapping (a paper can have more than one
affiliation/field associated with it).

Our goal now is to compare the explainability (as defined above) of these two
sets of features for the citation network. Since we want to compare them fairly,
we reduced the dataset to those nodes for which the dataset specifies both
features: that is, papers for which both the affiliations and the fields of
study are reported. In this way we obtained:

\begin{itemize}

    \item A graph $G=(N,A)$ where $N$ is a set of $18\,939\,155$ papers, and
    $A$ contains the $189\,465\,540$ citations between those papers.

    \item A set $F_a$ of $19\,834$ affiliations, and the association $\Z_a$
    between papers and affiliations. Each paper has between $1$ and $182$
    affiliations; on average, we have $1.36$ affiliations per paper.

    \item A set $F_f$ of $47\,269$ fields, and the association $\Z_f$ between
    papers and those fields of study. Each paper involves between $1$ and $200$
    fields; on average, we have $3.88$ fields per paper.

    \item As a further type of test, we performed the experiments also on the
    union $F_a \cup F_f$.

\end{itemize}

We proceeded then to evaluate the explainability of $F_a$ and $F_f$ for $G$ with
the same approach presented in \sec{sec:evaluating-lp}:

\begin{enumerate}
    \item We divide the set $N$ in ten folds $N_0, \dots, N_9$.
    \item For each fold $N_i$:
    \begin{enumerate}
        \item We apply Algorithm~\ref{alg:our-pa-alg} to the part of $A$ and
        $\Z$ related to the training set $\cup_{j \neq i}{N_j}$.
        \item We obtain a matrix $\W$.
        \item We compute the scores of our model with $\W$ on the test set $N_i$.
        \item We measure the precision-recall curve for these scores.
    \end{enumerate}
\end{enumerate}

In order to validate on real data the results we obtained in
\sec{sec:evaluating-lp} for synthetic data, we also carried out the same
procedure also with the $\W$ matrix found by \textsc{Naive}. As a result, we
obtained two ten-folded precision-recall curves for each of the three set of
features considered: $F_a, F_f$ and $F_a \cup F_f$.

\newcommand{\explainsn}[2]{
	\begin{center}
		\begin{tabular}{cc}
			\includegraphics[width=0.5\columnwidth]{img/cit-#1-prec-rec-NaiveEstimator}
			&
			\includegraphics[width=0.5\columnwidth]{img/cit-#1-prec-rec-LlamaEstimator} \\
            \textsc{Naive} & \textsc{Llama}
		\end{tabular}
		\caption{\label{fig:#1} Precision-recall curves of the Naive baseline and of
		\textsc{Llama}, when explaining the citation network using #2 as features.
        Different colors represent different folds used in cross-validation. }
    \end{center}
}

\begin{figure}[t]
    \explainsn{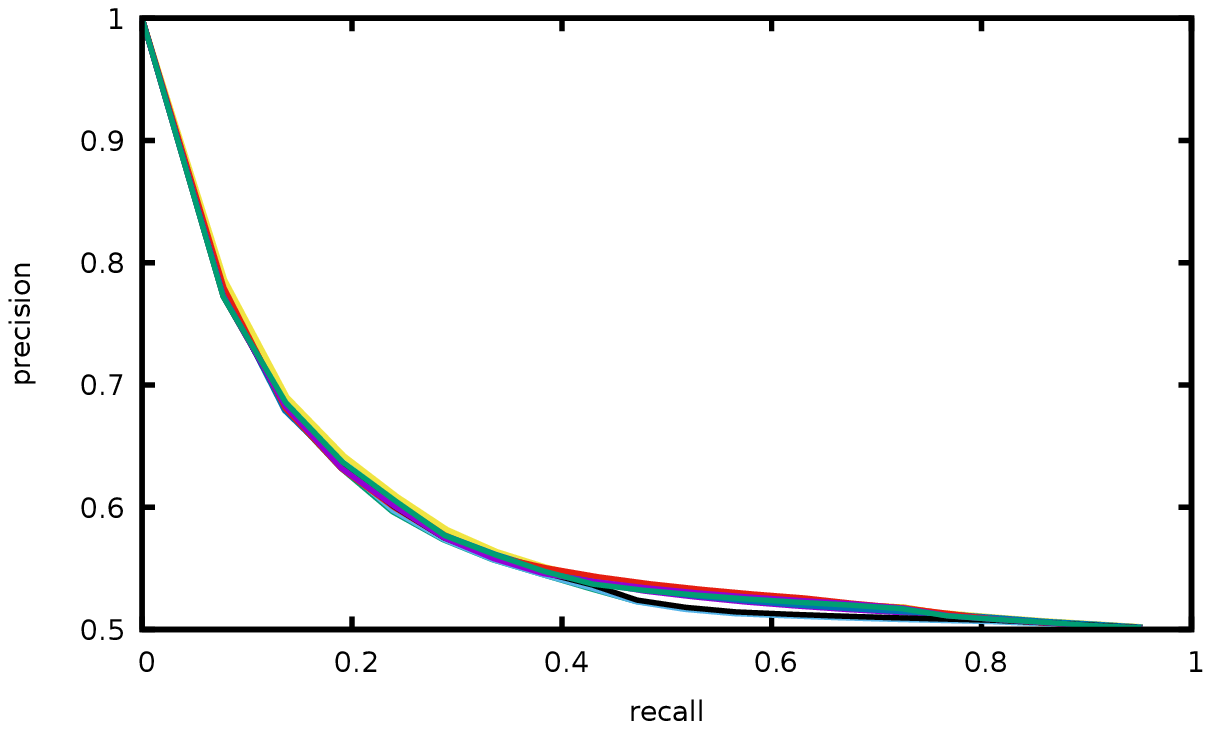}{the affiliation of authors}
    \explainsn{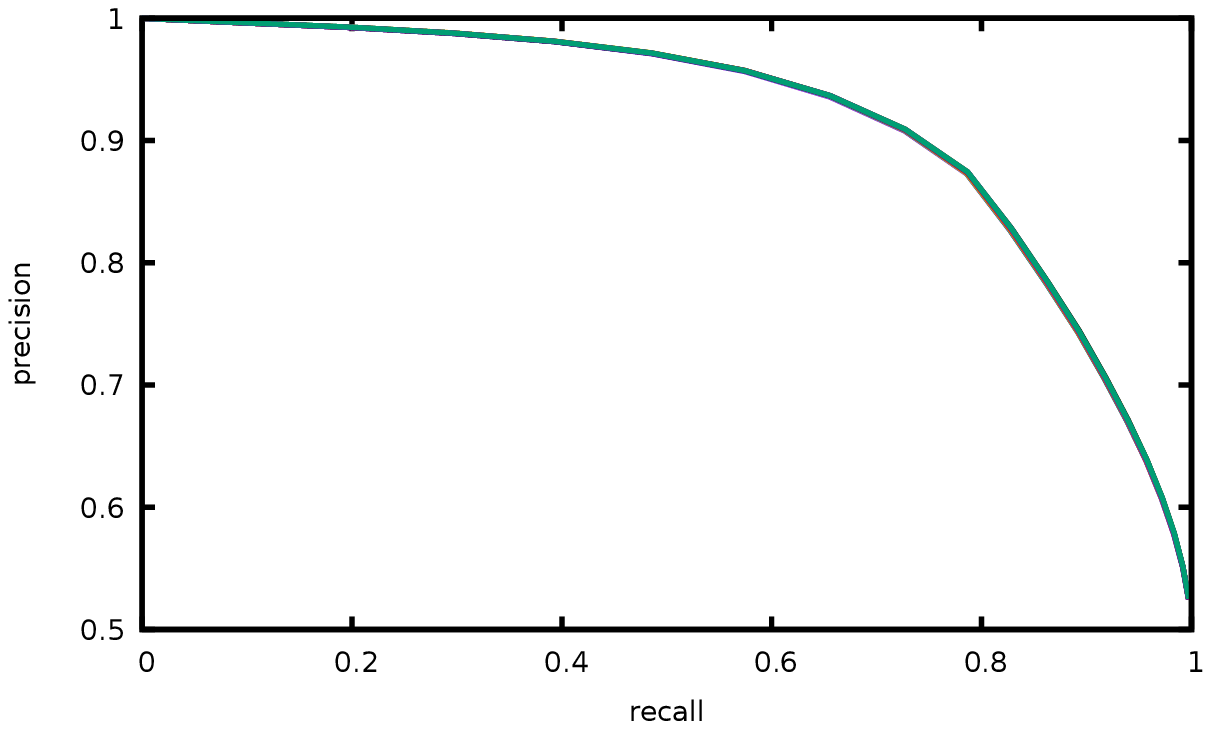}{the fields of study of each paper}
\end{figure}

\begin{table}[tb]
    \begin{center}
    \begin{tabular}{ llll  }
        \toprule
         &  Affiliations & Fields of study & Both \\
       \midrule
           	\textsc{Llama}
            & $\mathbf{.5551} \pm .0028$
            & $\mathbf{.9162} \pm .0003$
            & $.9210 \pm .0012$
           \\ \textsc{Llama} (natural order)
            & $.5446 \pm .0013$
            & $.9063 \pm .0004$
            & $.9176 \pm .0002$
           \\ \textsc{Naive}
            & $.5237 \pm .0005$
            & $.6007 \pm .0004$
            & $.6345 \pm .0002$
            \\ \bottomrule
        \end{tabular}

    \caption{   \label{tab:explainability-aupr} Area under the precision-recall
    curve of the Naive baseline and of \textsc{Llama}. For each of the
    feature sets considered, we report the mean and the standard deviation
    across the ten folds. We highlighted the \emph{explainability} for the
    citation network of the affiliations and of the fields of study,
    respectively.}
\end{center}
\end{table}

Furthermore, we are comparing two different orderings for node sequences in
\textsc{Llama}: one is purely random (the one we suggested in
Section~\ref{sec:pa-llama}), while the other is the natural order of nodes
in this case, i.e., the chronological order of paper publication. Please note,
however, that the 10-fold cross-validation is still operated a random (each train-test
split is performed randomly, regardless of ordering).

\subsection{Results}

In Table~\ref{tab:explainability-aupr} we report the explainability we obtained
(measured as the area under the precision-recall curves shown). We report in
Figure~\ref{fig:affiliations},~\ref{fig:fields}~and~\ref{fig:fields-affiliations}
the precision-recall curves for \textsc{Naive} and for \textsc{Llama}
concerning the feature set $F_a$, $F_f$ and $F_a \cup F_f$, respectively.

From the table, we can see that the explainability of the fields of study for
the citation network is much larger than that of the authors affiliations: the
first is above $92\%$, while the second is $56\%$. In this sense, our model
allows us to say that the fields of study of a paper explain very well its citations, while
the affiliations of its authors do not. This might not come as a surprise (the
relationship between the fields a paper belongs to and its citations is quite
natural) but our contribution here is the formal framework which allows us to
back this assertion with solid numbers, through \eqref{eq:basic-model} and
Algorithm~\ref{alg:our-pa-alg}.

We can further validate this statement by
looking at the explainability for $F_a \cup F_f$: its value of
$92.1\%$ is just faintly over the value of $91.6\%$ obtained for fields alone,
implying that the gain obtained by including the whole new set $F_a$ of
$19\,834$ features is practically negligible.

Finally, it is worth noting that the ordering of the node does not affect much
the results, that go from $91\%$ of the usual random order to $90\%$ for the
natural order.

\begin{figure}[t]
    \explainsn{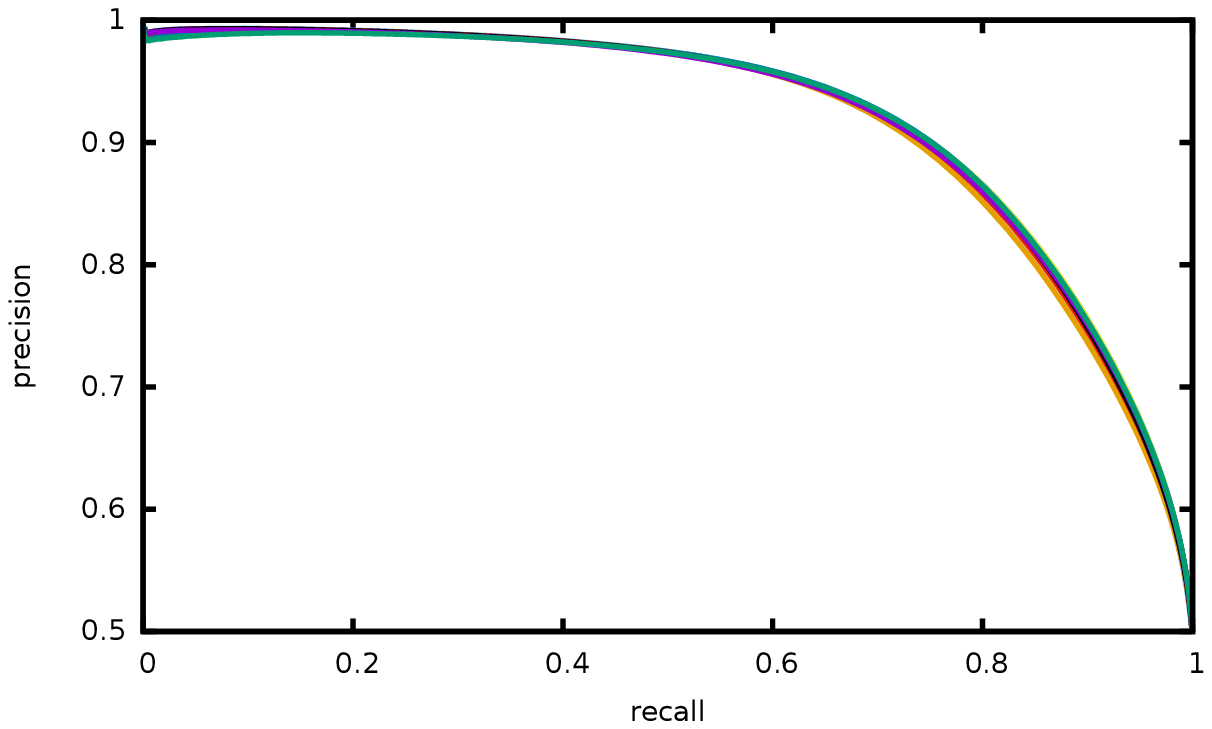}{both the affiliations and the fields, together,}
\end{figure}

We can grasp more details by looking at the specific precision-recall
curves. By comparing the \textsc{Llama} curve for affiliations in
Figure~\ref{fig:affiliations} and the one for fields in Figure~\ref{fig:fields},
we can see immediately that the latter depicts a valid classification instrument;
there, the precision/recall break-even point is around $83\%$. Also, we can
see some specific characteristic of the affiliation feature set: it is in fact
able to reach a large precision, but only in the very low range of recall. Here,
a precision of $83\%$ is possible only with a recall lower than $7\%$: the reason behind this
is that an author's affiliation is effective in encouraging a citation in a
very limited set of circumstances; we can conjecture that homophily within small
institutions could be an example.

Finally, let us remark how the results we obtained on synthetic data in
\sec{sec:evaluating-lp} are fully confirmed by the real data we presented here:
\textsc{Llama}, in all the three cases, behaves much better than \textsc{Naive}.
This is especially true for the feature set that actually explains the network:
for the fields of study, \textsc{Llama} is able to get a $91\%$ value for AUPR,
while the $\W$ matrix found by Naive approach can barely get a $63\%$. In
particular, precision-recall curves look similar to the one shown in
Figure~\ref{fig:sigmoid-bernoulli}, corresponding to the simulation obtained
with $\phi$ set to a sigmoid and $\W$ having a Bernoulli distribution; real data
is actually less shaky, due to the fact that we have $18$ millions of nodes
instead of the $10\,000$ used in the simulation. Besides confirming the validity
of \textsc{Llama}, this observation also confirms the goodness of our model in
explaining a real graph.

%

\section{Conclusions and future work}

In this work, we investigated large, feature-rich complex networks (networks
where each node is characterized by set of features).
Specifically, we wanted to analyze a model where node features induce the
formation of the links we observe. This hypothesis is reasonable in many
scenarios (the citation networks used in our experiments are just one example).
As discussed in~\sec{chap:state-of-the-art}, we employed the
Miller-Griffiths-Jordan model as our starting point. The problem we dealt with
was how to infer the latent feature-feature matrix: this matrix is the main
\emph{unknown} of the model; it determines how features interact between each
other to give raise to the observed links.

Specifically, we focused on the following scenario: assume to have complete
knowledge of a node-feature association matrix -- i.e., to know for every node,
the features it exhibits (embodied in the binary matrix $\Z$); also, assume to
have an (at least partial) knowledge of the links between these nodes (the
graph $G$).
Our goal was, given these elements,
to find the latent interaction between features that governs link formation in
the graph $G$; i.e., to discover the latent matrix $\W$ of our
model~\eqref{eq:basic-model}. This estimate alone allows us to use our model as
a possible way to predict which pair of nodes form a link. Other possible applications
include dimensionality reduction of the features, measuring semantic distance,
discovering hidden relationships, and so on.

While many possible methods are available in literature to attack these
problems, they generally only can handle small/medium sized networks, while we
are interested in large-scale networks. This ruled out many well-known
techniques, like MCMC. Our first approach was guided by a Naive Bayes scheme: we
demonstrated that a very simple equation to estimate the matrix can be derived
by assuming (naively) independence between features, and by making a few
assumptions to restrict our model. However, we pointed out how its naive
assumptions can cause problems in practical applications, and for this reason we
described a more sophisticated approach, based on perceptrons.

To link it formally with our model of choice, we assumed it to be deterministic
by choosing a step activation function $\phi$ in \eqref{eq:basic-model}.
This assumption allowed
us to align our model equation to a perceptron decision rule, by applying an
outer product kernel to the binary vectors $\z_{i}$ and $\z_{j}$ representing
the features in nodes $i$ and $j$, and to make the perceptron predictions
represent whether they form a link or not. In this way, the internal state of
the perceptron converges to the latent feature-feature matrix $\W$. We described
this learning-based approach, and analyzed what a classical bound on the number
of errors of a perceptron means in this case. Then, since any perceptron-like
algorithm can be adapted for this purpose, we chose the simple and fast
Passive-Aggressive algorithm~\cite{CDKSSOPAA} to concretely implement this
approach (Algorithm~\ref{alg:our-pa-alg}).

In the experimental section, we tested how this algorithm behaves on synthetic
data.
We generated graphs and node-feature associations according to the model presented
in~\cite{bolcrimon}, under different assumptions. In measuring the
outcomes, we adopted the same techniques as suggested
in~\cite{yang2015evaluating}:
specifically we measured the link prediction capability of the estimated $\W$
through a ten-fold cross-validation. Results showed how our learning approach
outperforms the Naive baseline in all the analyzed cases, except for the $\exp$
activation function.

Finally, we conducted an experiment on a real dataset, a citation network
composed by $18\,939\,155$ nodes and $189\,465\,540$ link; running the algorithm
required about $20$ minutes. In fact, we used the tools we developed for
estimating the feature-feature matrix in order to validate their performance on
real data, and to show how they can be used to assess which feature set can be
more useful in explaining the links of a network.

\bigskip

In this work, our main contribution consisted in laying out a bridge between
perceptron-like learning algorithms and feature-rich graph models; we formally
presented the connection between them, and we showed how they can be valuable
from a practical point of view when analyzing graphs that have tens of millions
of nodes or more.

We hope that the intersection of machine learning and complex network models
will attract more research in the future; many questions are left open on these
topics. Given a specific graph (possibly with features) how can we understand
what is the best model that can explain its links? Can this model also offer a
learning algorithm that allows us to make predictions about unknown nodes? A
full answer to this question would look, from one side, like a network ``family
tree'': it would enumerate possible models of networks by describing the
formation of their links, each being more or less reasonable depending on the
specific network at hand. From the other side, such a ``family tree'' would look
like a toolbox in the hands of the network scientist: each model should offer
algorithms for link prediction that could be more or less accurate or
computationally efficient.

Regarding the efficiency of algorithms for our models, there are some
alternatives that are left unexplored: for example other online algorithms, like
PEGASOS; also, we would like to investigate better formal connections between
neural networks and complex networks; for example, can deeper neural networks
also be read as a sensible feature-rich graph model?

Other future directions stem, on the contrary, from modifying our model. The
latent matrix $\W$, as reconstructed by the algorithms described in this paper,
will be dense; what happens if we reduce its density (e.g., by thresholding the
absolute value of its entries)? How much would that impact on our ability to
reconstruct $\A$? This density/precision tradeoff can be taken into consideration
from start: we may want to try to construct a latent matrix that satisfies some
constraints (e.g., on its density, or on its norm). This constrained version of
the problem may shed new light on the relation between features and links, and
can be a fruitful research direction.

Finally, we remark how it would be definitely important to test the proposed
techniques on other real feature-rich complex networks, in order to see in which
concrete cases they can improve over the current techniques for link prediction
and, more generally, for understanding hidden patterns in network data.

We consider these questions of primary importance, in order to be able to avoid
viewing graph mining algorithms as black boxes, but considering instead
what they could say about the structure and the evolution of specific complex
networks.

\bibliography{biblio/from-my-articles,biblio/tesi}

\begin{bibdiv}
\begin{biblist}

\bib{mixed-membership}{article}{
      author={Airoldi, Edoardo~M.},
      author={Blei, David~M.},
      author={Fienberg, Stephen~E.},
      author={Xing, Eric~P.},
       title={Mixed membership stochastic blockmodels},
        date={2008-06},
     journal={J. Mach. Learn. Res.},
      volume={9},
       pages={1981\ndash 2014},
}

\bib{aral1999sexual}{article}{
      author={Aral, S.~O},
      author={Hughes, J.P.},
      author={Stoner, B.},
      author={Whittington, W.},
      author={Handsfield, H.H.},
      author={Anderson, R.M.},
      author={Holmes, K.K.},
       title={Sexual mixing patterns in the spread of gonococcal and chlamydial
  infections.},
        date={1999},
     journal={American Journal of Public Health},
      volume={89},
      number={6},
       pages={825\ndash 833},
}

\bib{homophilymedia}{article}{
      author={Bisgin, Halil},
      author={Agarwal, Nitin},
      author={Xu, Xiaowei},
       title={A study of homophily on social media},
        date={2012},
     journal={World Wide Web},
      volume={15},
      number={2},
       pages={213\ndash 232},
}

\bib{BishopPatternRec}{book}{
      author={Bishop, C.M.},
       title={Pattern recognition and machine learning (information science and
  statistics)},
   publisher={Springer-Verlag New York, Inc.},
        date={2006},
}

\bib{bolcrimon}{article}{
      author={Boldi, Paolo},
      author={Crimaldi, Irene},
      author={Monti, Corrado},
       title={A network model characterized by a latent attribute structure
  with competition},
        date={2016},
        ISSN={0020-0255},
     journal={Information Sciences},
       pages={\ndash },
}

\bib{breiger}{article}{
      author={Breiger, Ronald~L.},
       title={{The Duality of Persons and Groups}},
        date={1974},
     journal={Social Forces},
      volume={53},
      number={2},
       pages={181\ndash 190},
}

\bib{caldarelli2002scale}{article}{
      author={Caldarelli, Guido},
      author={Capocci, Andrea},
      author={De~Los~Rios, Paolo},
      author={Munoz, Miguel~A},
       title={Scale-free networks from varying vertex intrinsic fitness},
        date={2002},
     journal={Physical review letters},
      volume={89},
      number={25},
       pages={258702},
}

\bib{CARVALHO2006SINGLE}{inproceedings}{
      author={Carvalho, V.R.},
      author={Cohen, W.W.},
       title={Single-pass online learning: Performance, voting schemes and
  online feature selection},
organization={ACM},
        date={2006},
   booktitle={Proc. of the 12th acm sigkdd},
       pages={548\ndash 553},
}

\bib{cesa2004generalization}{article}{
      author={Cesa-Bianchi, Nicolo},
      author={Conconi, Alex},
      author={Gentile, Claudio},
       title={On the generalization ability of on-line learning algorithms},
        date={2004},
     journal={IEEE Transactions on Information Theory},
      volume={50},
      number={9},
       pages={2050\ndash 2057},
}

\bib{cesa2006prediction}{book}{
      author={Cesa-Bianchi, Nicolo},
      author={Lugosi, G{\'a}bor},
       title={Prediction, learning, and games},
   publisher={Cambridge university press},
        date={2006},
}

\bib{chang2009relational}{inproceedings}{
      author={Chang, J.},
      author={Blei, D.M.},
       title={Relational topic models for document networks},
        date={2009},
   booktitle={International conference on artificial intelligence and
  statistics},
       pages={81\ndash 88},
}

\bib{chein2008graph}{book}{
      author={Chein, Michel},
      author={Mugnier, Marie-Laure},
       title={Graph-based knowledge representation: computational foundations
  of conceptual graphs},
   publisher={Springer Science \& Business Media},
        date={2008},
}

\bib{CDKSSOPAA}{article}{
      author={Crammer, K.},
      author={Dekel, O.},
      author={Keshet, J.},
      author={Shalev-Shwartz, S.},
      author={Singer, Y.},
       title={Online passive-aggressive algorithms},
        date={2006},
     journal={J. Mach. Learn. Res.},
      volume={7},
       pages={551\ndash 585},
}

\bib{cristianini2000introduction}{book}{
      author={Cristianini, Nello},
      author={Shawe-Taylor, John},
       title={An introduction to support vector machines and other kernel-based
  learning methods},
   publisher={Cambridge university press},
        date={2000},
}

\bib{dalvi2007efficient}{article}{
      author={Dalvi, Nilesh},
      author={Suciu, Dan},
       title={Efficient query evaluation on probabilistic databases},
        date={2007},
     journal={The VLDB Journal},
      volume={16},
      number={4},
       pages={523\ndash 544},
}

\bib{davis2006relationship}{inproceedings}{
      author={Davis, Jesse},
      author={Goadrich, Mark},
       title={The relationship between precision-recall and roc curves},
organization={ACM},
        date={2006},
   booktitle={Proceedings of the 23rd international conference on machine
  learning},
       pages={233\ndash 240},
}

\bib{doppa2010learning}{inproceedings}{
      author={Doppa, Janardhan~Rao},
      author={Yu, Jun},
      author={Tadepalli, Prasad},
      author={Getoor, Lise},
       title={Learning algorithms for link prediction based on chance
  constraints},
organization={Springer},
        date={2010},
   booktitle={Joint european conference on machine learning and knowledge
  discovery in databases},
       pages={344\ndash 360},
}

\bib{everett2013introduction}{book}{
      author={Everett, B},
       title={An introduction to latent variable models},
   publisher={Springer Science \& Business Media},
        date={2013},
}

\bib{GenALMA}{article}{
      author={Gentile, C.},
       title={A new approximate maximal margin classification algorithm},
        date={2002},
     journal={J. Mach. Learn. Res.},
      volume={2},
       pages={213\ndash 242},
}

\bib{MCMC-ML}{book}{
      author={Geyer, Charles~J.},
      author={Statistics), Minnesota Univ. (Minneapolis School~Of},
       title={{Markov Chain Monte Carlo Maximum Likelihood.}},
   publisher={Defense Technical Information Center},
        date={1992},
}

\bib{gong2012evolution}{inproceedings}{
      author={Gong, Neil~Zhenqiang},
      author={Xu, Wenchang},
      author={Huang, Ling},
      author={Mittal, Prateek},
      author={Stefanov, Emil},
      author={Sekar, Vyas},
      author={Song, Dawn},
       title={Evolution of social-attribute networks: measurements, modeling,
  and implications using google+},
organization={ACM},
        date={2012},
   booktitle={Proceedings of the 2012 acm conference on internet measurement
  conference},
       pages={131\ndash 144},
}

\bib{goodman1974exploratory}{article}{
      author={Goodman, Leo~A},
       title={Exploratory latent structure analysis using both identifiable and
  unidentifiable models},
        date={1974},
     journal={Biometrika},
      volume={61},
      number={2},
       pages={215\ndash 231},
}

\bib{GG06}{inproceedings}{
      author={Griffiths, Thomas~L.},
      author={Ghahramani, Zoubin},
       title={Infinite latent feature models and the indian buffet process},
        date={2005},
   booktitle={Advances in neural information processing systems},
   publisher={MIT Press},
       pages={475\ndash 482},
}

\bib{hall2009weka}{article}{
      author={Hall, Mark},
      author={Frank, Eibe},
      author={Holmes, Geoffrey},
      author={Pfahringer, Bernhard},
      author={Reutemann, Peter},
      author={Witten, Ian~H},
       title={The weka data mining software: an update},
        date={2009},
     journal={ACM SIGKDD explorations newsletter},
      volume={11},
      number={1},
       pages={10\ndash 18},
}

\bib{henderson2009applying}{inproceedings}{
      author={Henderson, K.},
      author={Eliassi-Rad, T.},
       title={Applying latent dirichlet allocation to group discovery in large
  graphs},
organization={ACM},
        date={2009},
   booktitle={Proc. 2009 acm symposium on applied computing},
       pages={1456\ndash 1461},
}

\bib{henry1983latent}{article}{
      author={Henry, Neil~W},
       title={Latent structure analysis},
        date={1983},
     journal={Encyclopedia of statistical sciences},
}

\bib{hens2009mining}{article}{
      author={Hens, N.},
      author={Goeyvaerts, N.},
      author={Aerts, M.},
      author={Shkedy, Z.},
      author={Van~Damme, P.},
      author={Beutels, P.},
       title={Mining social mixing patterns for infectious disease models based
  on a two-day population survey in belgium},
        date={2009},
     journal={BMC Infectious Diseases},
      volume={9},
      number={1},
       pages={5},
}

\bib{hertz1991introduction}{book}{
      author={Hertz, John},
      author={Krogh, Anders},
      author={Palmer, Richard~G},
       title={Introduction to the theory of neural computation},
   publisher={Basic Books},
        date={1991},
      volume={1},
}

\bib{latent-factor-models}{article}{
      author={Hoff, P.D.},
       title={Multiplicative latent factor models for description and
  prediction of social networks.},
        date={2009},
     journal={Computational and Mathematical Organization Theory},
      volume={15},
      number={4},
       pages={261\ndash 272},
}

\bib{hofman2008bayesian}{article}{
      author={Hofman, Jake~M},
      author={Wiggins, Chris~H},
       title={Bayesian approach to network modularity},
        date={2008},
     journal={Physical review letters},
      volume={100},
      number={25},
       pages={258701},
}

\bib{isella2011close}{article}{
      author={Isella, Lorenzo},
      author={Romano, Mariateresa},
      author={Barrat, Alain},
      author={Cattuto, Ciro},
      author={Colizza, Vittoria},
      author={Van~den Broeck, Wouter},
      author={others},
       title={Close encounters in a pediatric ward: measuring face-to-face
  proximity and mixing patterns with wearable sensors},
        date={2011},
     journal={PloS one},
      volume={6},
      number={2},
       pages={e17144},
}

\bib{JAPKOWICZ2002CLASSIMBALANCE}{article}{
      author={Japkowicz, N.},
      author={Stephen, S.},
       title={The class imbalance problem: A systematic study},
        date={2002},
     journal={Intelligent data analysis},
      volume={6},
      number={5},
       pages={429\ndash 449},
}

\bib{kemp2006learning}{inproceedings}{
      author={Kemp, Charles},
      author={Tenenbaum, Joshua~B},
      author={Griffiths, Thomas~L},
      author={Yamada, Takeshi},
      author={Ueda, Naonori},
       title={Learning systems of concepts with an infinite relational model},
        date={2006},
   booktitle={Aaai},
      volume={3},
       pages={5},
}

\bib{khan2015uncertain}{article}{
      author={Khan, Arijit},
      author={Chen, Lei},
       title={On uncertain graphs modeling and queries},
        date={2015},
     journal={Proceedings of the VLDB Endowment},
      volume={8},
      number={12},
       pages={2042\ndash 2043},
}

\bib{kim2011modeling}{article}{
      author={Kim, Myunghwan},
      author={Leskovec, Jure},
       title={Modeling social networks with node attributes using the
  multiplicative attribute graph model},
        date={2011},
     journal={arXiv preprint arXiv:1106.5053},
}

\bib{leskovec-mag}{article}{
      author={Kim, Myunghwan},
      author={Leskovec, Jure},
       title={Multiplicative attribute graph model of real-world networks},
        date={2012},
     journal={Internet Mathematics},
      volume={8},
      number={1-2},
       pages={113\ndash 160},
}

\bib{kim2013nonparametric}{inproceedings}{
      author={Kim, Myunghwan},
      author={Leskovec, Jure},
       title={Nonparametric multi-group membership model for dynamic networks},
        date={2013},
   booktitle={Advances in neural information processing systems},
       pages={1385\ndash 1393},
}

\bib{lattanzi_affiliation}{inproceedings}{
      author={Lattanzi, Silvio},
      author={Sivakumar, D.},
       title={Affiliation networks},
        date={2009},
   booktitle={Proc. of the forty-first annual acm symposium on theory of
  computing},
      series={STOC '09},
   publisher={ACM},
     address={New York, NY, USA},
       pages={427\ndash 434},
}

\bib{lazarsfeld1959latent}{article}{
      author={Lazarsfeld, Paul~F},
       title={Latent structure analysis},
        date={1959},
     journal={Psychology: A study of a science},
      volume={3},
       pages={476\ndash 543},
}

\bib{leskovec2010kronecker}{article}{
      author={Leskovec, Jure},
      author={Chakrabarti, Deepayan},
      author={Kleinberg, Jon},
      author={Faloutsos, Christos},
      author={Ghahramani, Zoubin},
       title={Kronecker graphs: An approach to modeling networks},
        date={2010},
     journal={Journal of Machine Learning Research},
      volume={11},
      number={Feb},
       pages={985\ndash 1042},
}

\bib{liu2009topic}{inproceedings}{
      author={Liu, Y.},
      author={Niculescu-Mizil, A.},
      author={Gryc, W.},
       title={Topic-link lda: joint models of topic and author community},
organization={ACM},
        date={2009},
   booktitle={Proc. 26th annual international conference on machine learning},
       pages={665\ndash 672},
}

\bib{mcpherson2001birds}{article}{
      author={McPherson, Miller},
      author={Smith-Lovin, Lynn},
      author={Cook, James~M},
       title={Birds of a feather: Homophily in social networks},
        date={2001},
     journal={Annual Review of Sociology},
      volume={27},
      number={1},
       pages={415\ndash 444},
}

\bib{mcsherry2008computing}{inproceedings}{
      author={McSherry, Frank},
      author={Najork, Marc},
       title={Computing information retrieval performance measures efficiently
  in the presence of tied scores},
organization={Springer},
        date={2008},
   booktitle={European conference on information retrieval},
       pages={414\ndash 421},
}

\bib{meeds2006modeling}{inproceedings}{
      author={Meeds, Edward},
      author={Ghahramani, Zoubin},
      author={Neal, Radford~M},
      author={Roweis, Sam~T},
       title={Modeling dyadic data with binary latent factors},
        date={2006},
   booktitle={Advances in neural information processing systems},
       pages={977\ndash 984},
}

\bib{menche2015uncovering}{article}{
      author={Menche, J{\"o}rg},
      author={Sharma, Amitabh},
      author={Kitsak, Maksim},
      author={Ghiassian, Susan~Dina},
      author={Vidal, Marc},
      author={Loscalzo, Joseph},
      author={Barab{\'a}si, Albert-L{\'a}szl{\'o}},
       title={Uncovering disease-disease relationships through the incomplete
  interactome},
        date={2015},
     journal={Science},
      volume={347},
      number={6224},
       pages={1257601},
}

\bib{menon2011link}{inproceedings}{
      author={Menon, Aditya~Krishna},
      author={Elkan, Charles},
       title={Link prediction via matrix factorization},
organization={Springer},
        date={2011},
   booktitle={Joint european conference on machine learning and knowledge
  discovery in databases},
       pages={437\ndash 452},
}

\bib{MGJ2009}{inproceedings}{
      author={Miller, K.T.},
      author={Griffiths, T.L.},
      author={Jordan, M.I.},
       title={Nonparametric latent feature models for link prediction.},
        date={2009},
   booktitle={In nips},
      editor={Bengio, Y.},
      editor={Schuurmans, D.},
      editor={Lafferty, J.D.},
      editor={Williams, C.K.~I.},
      editor={Culotta, A.},
   publisher={Curran Associates, Inc.},
       pages={1276\ndash 1284},
}

\bib{MRZZAC}{inproceedings}{
      author={Monti, Corrado},
      author={Rozza, A.},
      author={Zappella, G.},
      author={Zignani, M.},
      author={Arvidsson, A.},
      author={Colleoni, E.},
       title={Modelling political disaffection from twitter data},
organization={ACM},
        date={2013},
   booktitle={Proc. of the 2nd int. wisdom},
       pages={3},
}

\bib{mossong2008social}{article}{
      author={Mossong, J.},
      author={Hens, N.},
      author={Jit, M.},
      author={others},
       title={Social contacts and mixing patterns relevant to the spread of
  infectious diseases},
        date={2008},
     journal={PLoS Med},
      volume={5},
      number={3},
       pages={e74},
}

\bib{nowicki2001estimation}{article}{
      author={Nowicki, Krzysztof},
      author={Snijders, Tom A~B},
       title={Estimation and prediction for stochastic blockstructures},
        date={2001},
     journal={Journal of the American Statistical Association},
      volume={96},
      number={455},
       pages={1077\ndash 1087},
}

\bib{pfeiffer2014attributed}{inproceedings}{
      author={Pfeiffer~III, Joseph~J},
      author={Moreno, Sebastian},
      author={La~Fond, Timothy},
      author={Neville, Jennifer},
      author={Gallagher, Brian},
       title={Attributed graph models: Modeling network structure with
  correlated attributes},
organization={ACM},
        date={2014},
   booktitle={Proceedings of the 23rd international conference on world wide
  web},
       pages={831\ndash 842},
}

\bib{potamias2010k}{article}{
      author={Potamias, Michalis},
      author={Bonchi, Francesco},
      author={Gionis, Aristides},
      author={Kollios, George},
       title={K-nearest neighbors in uncertain graphs},
        date={2010},
     journal={Proceedings of the VLDB Endowment},
      volume={3},
      number={1-2},
       pages={997\ndash 1008},
}

\bib{rosenblatt1958perceptron}{article}{
      author={Rosenblatt, Frank},
       title={The perceptron: a probabilistic model for information storage and
  organization in the brain.},
        date={1958},
     journal={Psychological review},
      volume={65},
      number={6},
       pages={386},
}

\bib{rosenblatt1961principles}{techreport}{
      author={Rosenblatt, Frank},
       title={Principles of neurodynamics. perceptrons and the theory of brain
  mechanisms},
 institution={DTIC Document},
        date={1961},
}

\bib{russellnorvig}{misc}{
      author={Russell, Stuart},
      author={Norvig, Peter},
       title={Artificial intelligence: A modern approach. 2010},
   publisher={Prentice Hall},
        date={2010},
}

\bib{sculley2007online}{inproceedings}{
      author={Sculley, D},
       title={Online active learning methods for fast label-efficient spam
  filtering.},
        date={2007},
   booktitle={Ceas},
      volume={7},
       pages={143},
}

\bib{sinha2015overview}{inproceedings}{
      author={Sinha, Arnab},
      author={Shen, Zhihong},
      author={Song, Yang},
      author={Ma, Hao},
      author={Eide, Darrin},
      author={Hsu, Bo-june~Paul},
      author={Wang, Kuansan},
       title={An overview of microsoft academic service (mas) and
  applications},
organization={ACM},
        date={2015},
   booktitle={Proceedings of the 24th international conference on world wide
  web},
       pages={243\ndash 246},
}

\bib{blockmodel}{article}{
      author={Snijders, Tom~A.B.},
      author={Nowicki, Krzysztof},
       title={Estimation and prediction for stochastic blockmodels for graphs
  with latent block structure},
        date={1997},
        ISSN={0176-4268},
     journal={Journal of Classification},
      volume={14},
      number={1},
       pages={75\ndash 100},
}

\bib{stouffer1950measurement}{article}{
      author={Stouffer, Samuel~A},
      author={Guttman, Louis},
      author={Suchman, Edward~A},
      author={Lazarsfeld, Paul~F},
      author={Star, Shirley~A},
      author={Clausen, John~A},
       title={Measurement and prediction.},
        date={1950},
}

\bib{trefethen1997numerical}{book}{
      author={Trefethen, Lloyd~N},
      author={Bau~III, David},
       title={Numerical linear algebra},
   publisher={Siam},
        date={1997},
      volume={50},
}

\bib{wu2008interpreting}{article}{
      author={Wu, Ho~Chung},
      author={Luk, Robert Wing~Pong},
      author={Wong, Kam~Fai},
      author={Kwok, Kui~Lam},
       title={Interpreting tf-idf term weights as making relevance decisions},
        date={2008},
     journal={ACM Transactions on Information Systems (TOIS)},
      volume={26},
      number={3},
       pages={13},
}

\bib{xu2006learning}{article}{
      author={Xu, Zhao},
      author={Tresp, Volker},
      author={Yu, Kai},
      author={Kriegel, Hans-Peter},
       title={Learning infinite hidden relational models},
        date={2006},
     journal={Uncertainity in Artificial Intelligence (UAI2006)},
}

\bib{yang2015evaluating}{article}{
      author={Yang, Yang},
      author={Lichtenwalter, Ryan~N},
      author={Chawla, Nitesh~V},
       title={Evaluating link prediction methods},
        date={2015},
     journal={Knowledge and Information Systems},
      volume={45},
      number={3},
       pages={751\ndash 782},
}

\end{biblist}
\end{bibdiv}

\end{document}